\documentclass[%
 reprint,
 superscriptaddress,
 amsmath,amssymb,
 aps,
 prl,
]{revtex4-2}

\usepackage{bm}
\usepackage[hidelinks]{hyperref}
\usepackage{cleveref}
\usepackage{csquotes}
\usepackage{graphicx}
\usepackage{outlines}
\usepackage{bbold}
\usepackage{siunitx}
\usepackage{makecell}
\usepackage{braket}
\usepackage{amsthm}

\renewcommand{\vec}[1]{\bm{#1}}
\newcommand{\abs}[1]{\left|#1\right|}
\newcommand{\E}[1]{\left\langle #1\right\rangle}

\newcommand{\ee}{\mathrm{e}}
\newcommand{\ii}{\mathrm{i}}

\newcommand{\widebar}{\overline}

\newtheorem{theorem}{Theorem}
\newtheorem{lemma}{Lemma}

\begin{document}

\title{Visualizing two qubits}

\author{Simon Grosse-Holz}\email{s.grosse.holz@alum.mit.edu}

\date{\today}

\begin{abstract}
	I propose a simple visualization for pure two-qubit states, constructed
	from three components: two Bloch balls for the marginal qubits (the
	mixed states obtained by ignoring the other qubit, respectively) and an
	entanglement component consisting of a rotoreflection and a
	contraction. I prove that this picture accurately captures the local
	and global structure of the relevant state space without gaps or
	degeneracies: the underlying parametrization is a global diffeomorphism
	onto the state space. Finally, I use this picture to visualize
	conditional states, the entanglement degree(s) of freedom, and single
	qubit rotations.
\end{abstract}

\maketitle

Visualizing a single qubit on the Bloch sphere
\cite{feynman_geometrical_1957,nielsen_quantum_2010} is one of the most
well-established pictures in quantum mechanics. It has provided generations of
students and researchers with intuitive understanding whenever two-level
quantum systems are studied: from the seminal Stern-Gerlach experiment
\cite{gerlach_experimentelle_1922} over NMR \cite{rabi_new_1938} and Bell's
inequalities \cite{bell_einstein_1964}, to (of course) large realms of Quantum
Information \cite{nielsen_quantum_2010}.  This immense explanatory power,
however, drops sharply as soon as another quintessentially quantum phenomenon
is concerned: entanglement.

Consider the case of just two qubits. To a classical mind, this means: two
Bloch spheres, so 4 degrees of freedom. But the true state space
$\mathbb{CP}^3$ (4 complex parameters, modulo normalization and global phase)
is 6-dimensional! It is thus immediately evident that the picture is missing
two degrees of freedom; which (essentially by definition) encode the
entanglement in the system \cite{kus_geometry_2001,mosseri_geometry_2001}. The
\enquote{na\"ive} two Bloch sphere picture is thus not wrong, but incomplete:
it captures exactly the separable states.

\begin{figure}
	\includegraphics{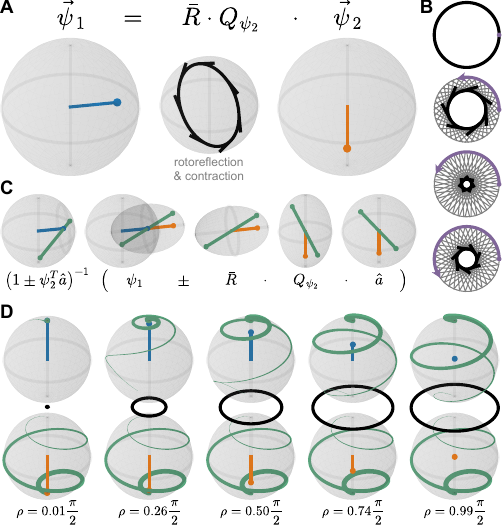}
	\caption{
Visualizing two-qubit states.
(A) A generic two-qubit state $\ket{\psi}$. The marginals $\vec\psi_1$ (blue,
left) and $\vec\psi_2$ (orange, right) are drawn in their respective Bloch
spheres, connected by the correlation matrix $\Gamma = \bar{R}Q_{\psi_2}$
through \cref{eq:obs:1}. We account for the contraction $Q_{\psi_2}$ by
rescaling the central sphere by the concurrence $\sqrt{1-\abs{\vec\psi_2}^2}$
and draw the rotoreflection $\bar{R}$ inside as detailed in (B).
(B) To visualize a rotoreflection, we identify its associated mirror plane
(here: the page) and the rotation angle $\omega$ within that plane (purple;
increasing top to bottom). We draw a circle in the mirror plane of radius
$\abs{\cos\frac{\omega}{2}}$, decorated with \enquote{flags} of length
$\frac{1}{2}\sin\omega$ indicating the direction of rotation. These choices
align with the geometry of the straight lines (gray) between points on the
unit circle and their image under the rotation; making the whole construction
reminiscent of a mechanical aperture stop.
(C) Constructing conditionals for a measurement along an axis $\hat{a}$
(green), following \cref{eq:conditional}. Right to left: the two possible
measurement outcomes define two points $\pm\hat{a}$ on the Bloch sphere of
qubit 2.  We contract by $Q_{\psi_2}$; rotoreflect by $\bar{R}$; add
$\vec\psi_1$; and normalize, to identify the state of qubit 1, conditional on
the outcome of the measurement of qubit 2.
(D) Same construction of conditional states as in (C), but for states of the
form $\cos\frac{\rho}{2}\ket{\uparrow\downarrow} +
\sin\frac{\rho}{2}\ket{\downarrow\uparrow}$ (allowing us to vary the degree of
entanglement; increasing left to right) and various potential measurement
outcomes for qubit 2 (bottom, green spiral). With little entanglement, the
conditional states become very insensitive to the measurement outcome on the
other qubit.
	}
	\label{fig:1}
\end{figure}

The full geometry of the two-qubit state space has been studied from various
angles around the turn of the millenium
\cite{brody_geometric_2001,kus_geometry_2001,mosseri_geometry_2001,bengtsson_cpn_2002,mosseri_two_2003,havel_bloch-sphere-type_2004}
with continued interest since then
\cite{gamel_entangled_2016,wie_two-qubit_2020,rau_symmetries_2021,filatov_towards_2024,filatov_roto-reflection_2026,shoji_visualization_2026}.
The point of the present paper is thus not just to propose yet another picture
(\cref{fig:1}A, kindred in spirit to \cite{filatov_roto-reflection_2026}), but
also to rigorously prove that it is in fact complete (\cref{thm:bijective}) and
faithful (\cref{thm:diffeo}); thus ensuring that the intuition derived from it
is \enquote{practically useful, because technically accurate}.

While the synthesis is novel to my knowledge, most building blocks for these
technical results are available in the literature: the rotoreflection geometry
established by \cref{thm:bijective} has been reported recently by Filatov and
Auzinsh \cite{filatov_roto-reflection_2026}, while the constant-concurrence
foliation and Hopf fibration
\cite{hopf_uber_1931,lyons_elementary_2003,treisman_young_2009} underlying the
smoothness results in \cref{thm:diffeo} already feature in
\cite{kus_geometry_2001,mosseri_two_2003}.

After developing the technical machinery and the picture derived from it, I
close by applying this intuition to joint measurements and conditional states;
the freedom \enquote{hidden} in the entanglement; and the two perspectives on
single qubit transformations ($SU(2)$ vs.\ $SO(3)$).

Consider a generic two-qubit state
\begin{equation}\label{eq:def:psi}
	\ket{\psi} \equiv a\ket{\uparrow\uparrow} + b\ket{\uparrow\downarrow} + c\ket{\downarrow\uparrow} + d\ket{\downarrow\downarrow}\,,
\end{equation}
where $a, b, c, d \in \mathbb{C}$ with $\abs{a}^2 + \abs{b}^2 + \abs{c}^2 +
\abs{d}^2 = 1$. Observables on this state are $4\times 4$ hermitian matrices
$M^\dagger = M$, with possible measurement outcomes equal to the eigenspectrum
of $M$ and expectation values given by $\E{M} \equiv \braket{\psi|M|\psi}$.
The observables $M$ form a 16-dimensional vector space over $\mathbb{R}$; a
basis of this space is given by tensor products of the three Pauli matrices
plus identity, $\left(\sigma_0, \vec\sigma\right) \equiv \left(\mathbb{1}_2,
\sigma_x, \sigma_y, \sigma_z\right)$. By linearity, the expected value of any
observable $M$ can be constructed from the expectations over this basis;
motivating us to define
\begin{align}
	\vec{\psi}_1 {}\equiv{}& \E{\vec\sigma\otimes\mathbb{1}_2}\,, &
	\vec{\psi}_2 {}\equiv{}& \E{\mathbb{1}_2\otimes\vec\sigma}\,, &
	\Gamma {}\equiv{}& \E{\vec\sigma\otimes\vec\sigma}\,,
\end{align}
while for the 16th basis observable
\begin{equation}
	\E{\mathbb{1}_2\otimes\mathbb{1}_2} = \abs{a}^2 + \abs{b}^2 + \abs{c}^2 + \abs{d}^2 = 1
\end{equation}
by state normalization. We display the expectation values for all 16 basis
observables in \cref{tab:expectations}.
\begin{table*}
\setlength{\tabcolsep}{8pt}
\setlength{\extrarowheight}{10pt}
\caption{
	Expectation values $E_{ij} \equiv
	\bra{\psi}\sigma_i\otimes\sigma_j\ket{\psi}$ for the state $\ket{\psi}$
	of \cref{eq:def:psi}.
}
\label{tab:expectations}
\begin{tabular}{c||c c c c}
	$\E{\sigma_i\otimes\sigma_j}$ & $\mathbb{1}$ & $\sigma_x$ & $\sigma_y$ & $\sigma_z$ \\[5pt]
	\hline\hline
	$\mathbb{1}$ & $\abs{a}^2 + \abs{b}^2 + \abs{c}^2 + \abs{d}^2$ & $2\Re\left(a^*b + c^*d\right)$ & $2\Im\left(a^*b+c^*d\right)$ & $\abs{a}^2 - \abs{b}^2 + \abs{c}^2 - \abs{d}^2$ \\
	$\sigma_x$ & $2\Re\left(a^*c+b^*d\right)$ & $2\Re\left(a^*d+b^*c\right)$ & $2\Im\left(a^*d-b^*c\right)$ & $2\Re\left(a^*c-b^*d\right)$ \\
	$\sigma_y$ & $2\Im\left(a^*c+b^*d\right)$ & $2\Im\left(a^*d+b^*c\right)$ & $-2\Re\left(a^*d-b^*c\right)$ & $2\Im\left(a^*c-b^*d\right)$ \\
	$\sigma_z$ & $\abs{a}^2 + \abs{b}^2 - \abs{c}^2 - \abs{d}^2$ & $2\Re\left(a^*b - c^*d\right)$ & $2\Im\left(a^*b-c^*d\right)$ & $\abs{a}^2 - \abs{b}^2 - \abs{c}^2 + \abs{d}^2$
\end{tabular}
\end{table*}

This matrix of basis expectations lies on a specific lowdimensional submanifold
of $\mathbb{R}^{4\times 4}$,
\begin{equation}
	\mathcal{E} \equiv \left\{ E\in\mathbb{R}^{4\times4} \;\middle|\; \exists\ket{\psi} : E_{ij} = \bra{\psi}\sigma_i\otimes\sigma_j\ket{\psi} \right\}\,,
\end{equation}
which is diffeomorphic to $\mathbb{CP}^3$ ($\mathbb{C}^4\setminus\{0\}$ modulo
normalization and global phase) by construction. From a physical point of view,
it is this 6-dimensional manifold of \enquote{observably different} states
whose structure we need to understand to gain intuition for the two-qubit
system.

To start building this understanding, consider
\begin{equation}\label{eq:decompE}
	E\equiv \begin{pmatrix} 1 & \vec\psi_2^T \\ \vec\psi_1 & \Gamma \end{pmatrix} \in\mathcal{E}\,.
\end{equation}
A few observations follow by direct calculation from \cref{tab:expectations}:
\begin{gather}
	\vec\psi_1 = \Gamma\vec\psi_2 \,, \label{eq:obs:1} \\
	\det\Gamma = -4\abs{ad-bc}^2 \equiv -\sin^2\rho\,, \label{eq:obs:2} \\
	\abs{\vec\psi_1}^2 = \abs{\vec\psi_2}^2 = 1 + \det\Gamma \equiv \cos^2\rho \,, \label{eq:obs:3} \\
	\Gamma = \vec\psi_1\otimes\vec\psi_2 \;\text{iff $\abs{\vec\psi_1} = \abs{\vec\psi_2} = 1$}\,. \label{eq:obs:4}
\end{gather}
where we introduce 
\begin{equation}
	\rho \equiv \arcsin\left(2\abs{ad-bc}\right) \in \left[0, \frac{\pi}{2}\right]\,.
\end{equation}
Note that $\sin\rho = 2\abs{ad-bc}$ is the \emph{concurrence} of the two-qubit
system \cite{hill_entanglement_1997} and between 0 and 1 by state
normalization, guaranteeing that $\rho$ is well-defined. Meanwhile $\cos\rho =
\abs{\vec\psi_1} = \abs{\vec\psi_2}$ gives the length of the marginals i.e.\
the \emph{purity} of the state $\ket{\psi}$.

In anticipation of the technical section of the paper, we define
\begin{align}
	O^-(3) {}\equiv{}& \left\{R\in O(3) \;\middle|\; \det R = -1\right\} \\
	{}={}& O(3)\setminus SO(3)\,,
\end{align}
the set of rotoreflections. Note that in 3D we can flip the determinant with a
global sign: $O^-(3)\ni\bar{R}\mapsto -\bar{R}\in SO(3)$ and thus pointwise
identify $O^-(3)\simeq SO(3)$. In theory, $O^-(3)$ then inherits $SO(3)$'s Lie
group structure; the group operation, however, would be matrix multiplication
with an extra global sign flip, such that $-\mathbb{1}_3\in O^-(3)$ is the
identity element. We do not use this group structure in the following and note
it only for curiosity. We do, however, use the smooth (sub-)manifold structure
of $O^-(3)\simeq SO(3) \subset\mathbb{R}^{3\times 3}$. Beyond $O^-(3)$, we use
standard notation, such as e.g.\ $\widebar{B^3} \equiv B^3\sqcup S^2$ for the
closed unit ball in $\mathbb{R}^3$ as the disjoint union of the open unit ball
and the sphere.

Thus equipped with some preliminary definitions and observations
\labelcref{eq:obs:1,eq:obs:2,eq:obs:3,eq:obs:4} about the space of observably
different two-qubit states $\mathcal{E}$, let us revisit the quest for its full
structure:

\begin{theorem}[see also \cite{filatov_roto-reflection_2026}]\label{thm:bijective}
	Let
	\begin{equation}
		\mathcal{P} \equiv O^-(3)\times B^3 \sqcup S^2\times S^2
	\end{equation}
	and $p\in\mathcal{P}$.
	\begin{itemize}
		\item For $p = \left(\bar{R}, \vec w\right) \in O^-(3)\times
			B^3$ (the first component), set
			\begin{align}
				\vec\psi_1 {}={}& \bar{R}\vec w\,, & \vec\psi_2 {}={}& \vec w\,,
			\end{align}
			and
			\begin{equation}
				\Gamma = \bar{R}\left(\frac{\vec w\otimes\vec w}{1+\sqrt{1-\abs{\vec w}^2}} + \sqrt{1-\abs{\vec w}^2}\right)\,.
			\end{equation}
		\item For $p = \left(\vec v, \vec w\right)\in S^2\times S^2$
			(the second component), set
			\begin{align}
				\vec\psi_1 {}={}& \vec v\,, & \vec\psi_2 {}={}& \vec w\,, & \Gamma {}={}& \vec v \otimes \vec w\,.
			\end{align}
	\end{itemize}
	Let now
	\begin{equation}
		E(p) \equiv \begin{pmatrix} 1 & \vec\psi_2^T \\ \vec\psi_1 & \Gamma \end{pmatrix}\,.
	\end{equation}
	Then $E(\mathcal{P}) = \mathcal{E}$ and the map $E:
	\mathcal{P}\to\mathcal{E}$ is bijective.

	\begin{proof}[Motivation (proof in appendix)]
		Any pure two-qubit state $\ket{\psi}$ is locally unitary
		equivalent \cite{kraus_local_2010} to
		\begin{equation}
			\ket{\psi_0} \equiv \cos\frac{\rho}{2}\ket{\uparrow\downarrow} + \sin\frac{\rho}{2}\ket{\downarrow\uparrow}\,,
		\end{equation}
		whose expectation values are easily identified from
		\cref{tab:expectations}.  Locally unitary transformations act
		on those expectation values as rotation matrices, i.e.\ given
		the generic state $\ket{\psi}$ there are $R_1,R_2\in SO(3)$
		such that its expectation values can be written as
		\begin{equation}
			\begin{pmatrix} 1 & \\ & R_1 \end{pmatrix} \begin{pmatrix} 1 & & & -\cos\rho \\ & \sin\rho & & \\ & & \sin\rho & \\ \cos\rho & & & -1 \end{pmatrix} \begin{pmatrix} 1 & \\ & R_2^T \end{pmatrix}\,.
		\end{equation}
		By decomposing these expectations as in \cref{eq:decompE} and
		\begin{equation}
			R_i \equiv R_z(\phi_i) R_y(\theta_i) R_z(\omega_i) \equiv S(\theta_i, \phi_i)R_z(\omega_i)
		\end{equation}
		into Euler angles, we identify
		\begin{align}
			\vec\psi_1 {}={}& \hat{s}(\theta_1, \phi_1)\cos\rho \\
			\vec\psi_2 {}={}& -\hat{s}(\theta_2, \phi_2)\cos\rho \\
			\Gamma {}={}& S(\theta_1, \phi_1) R_z(\omega_1 + \omega_2) P_z Q_z(\rho) S^T(\theta_2, \phi_2) \,,
		\end{align}
		where $\hat{s}(\theta, \phi) \equiv S(\theta, \phi)\hat{z}
		\in S^2$ is the unit vector with polar angle $\theta$ and
		azimuthal angle $\phi$, while $P_z \equiv \mathbb{1}_3 -
		2\hat{z}\otimes\hat{z}$ is the reflection along the $z$-axis
		and
		\begin{equation}
			Q_z(\rho) \equiv \sin\rho + \left(1-\sin\rho\right)\hat{z}\otimes\hat{z}
		\end{equation}
		a contraction in the $xy$-plane. Note that $Q_z$ and $R_z$
		commute, while $P_zR_z = R_z^TP_z$. The given construction of
		$E(p)$---and specifically
		$\Gamma\equiv\bar{R}Q_{\psi_2}(\rho)$---then follows by
		realizing that
		\begin{equation}
			S Q_z(\rho) S^T = \sin\rho + \left(1-\sin\rho\right)\hat{s}\otimes\hat{s} \equiv Q_s(\rho)
		\end{equation}
		and
		\begin{equation}
			S(\theta_1, \phi_1) R_z(\omega_1+\omega_2) P_z S^T(\theta_2, \phi_2) \equiv \bar{R} \in O^-(3)
		\end{equation}
		is a generic rotoreflection since $\bar{R}R_2P_z = R_1$ was
		arbitrary. For $\rho > 0$, $Q_s(\rho)$ is invertible, letting
		us identify $\bar{R}$ from $E(p)$; while $\bar{R}$ is
		underdetermined from $E(p)$ for separable states ($\rho = 0$),
		which are fully parametrized by the marginals
		$\vec\psi_1,\vec\psi_2\in S^2$. The disjoint construction of
		$\mathcal{P}$ in the \namecref{thm:bijective} thus ensures the
		existence of the inverse $E^{-1}:\mathcal{E} \to \mathcal{P}$.
	\end{proof}
\end{theorem}

It is a good excercise to check that \cref{eq:obs:1,eq:obs:2,eq:obs:3,eq:obs:4}
indeed hold for all $p\in\mathcal{P}$.

\begin{lemma}\label{thm:smooth}
	$\mathcal{P}$ with the quotient topology inherited from
	$O^-(3)\times\widebar{B^3}$ via the quotient map
	\begin{equation}
		q\left(\bar{R}, \vec w\right) \equiv \begin{cases}
			\left(\bar{R}, \vec w\right) & \text{if $\abs{\vec w} < 1$} \\
			\left(\bar{R}\vec w, \vec w\right) & \text{if $\abs{\vec w} = 1$,}
		\end{cases}
	\end{equation}
	is a smooth manifold without boundary.

	\begin{proof}[Motivation (proof in appendix)]
		Trivial for $\abs{\vec w} < 1$. For $\abs{\vec w} > 0$ we have
		\begin{equation}
			O^-(3)\times \widebar{B^3}\setminus\{0\} \simeq O^-(3)\times S^2\times (0, 1]\,.
		\end{equation}
		By Hopf fibration, $O^-(3)\times S^2$ is a $S^1$-fiber bundle
		over $S^2\times S^2$, allowing us to write the above as a
		$S^1\times (0, 1]$-fiber bundle over the same base. The
		quotient then maps this half-open cylindrical fiber to the open
		unit disc $B^2$ by identifying all points along the (single)
		edge. The natural smooth structure on the resulting $B^2$-fiber
		bundle is compatible with the one on $O^-(3)\times B^3$, such
		that now we have two smooth manifolds that cover $\mathcal{P}$
		and are diffeomorphic on their overlap; thus enabling the
		construction of a smooth atlas on all of $\mathcal{P}$.
	\end{proof}
\end{lemma}

\begin{theorem}\label{thm:diffeo}
	The map $E:\mathcal{P}\to\mathcal{E}$ of \cref{thm:bijective} is a
	diffeomorphism.

	\begin{proof}[Motivation (proof in appendix)]
		Bijectivity was shown in \cref{thm:bijective}, so we only need
		smoothness of $E$ and $E^{-1}$; both are trivial for
		$\abs{\vec{w}} < 1$. For $\abs{\vec{w}} > 0$ rewrite $E(p)$ in
		terms of the $B^2$-fiber bundle constructed in
		\cref{thm:smooth}. The non-smooth expressions $\abs{z}$ and
		$\arg z$ in the bundle coordinate $z\in B^2\subset\mathbb{C}$
		recombine into $\Re z = \abs{z}\cos\arg z$ and $\Im z =
		\abs{z}\sin\arg z$, making the construction smooth in
		$\left(\Re z, \Im z\right)^T\in B^2\subset\mathbb{R}^2$.
	\end{proof}
\end{theorem}

\Cref{fig:1}A employs the parametrization of \cref{thm:bijective} to propose a
visualization of pure two-qubit states (the construction for depicting a
rotoreflection $\bar{R}$ is illustrated in \cref{fig:1}B).  While
\cref{thm:bijective} guarantees the pointwise accuracy of this picture,
\cref{thm:diffeo} assures us that the local and global structure of the state
space $\mathcal{E}$ is indeed correctly represented. The central sphere is
globally scaled by the concurrence and thus vanishes for separable states;
leaving in this case exactly the \enquote{na\"ive} two Bloch spheres for
unentangled qubits.

As a direct application of this picture, let us visualize the effect of
entanglement in \cref{fig:1}C,D: what do we learn about qubit 1 by measuring
qubit 2? Consider a measurement of qubit 2 along an axis $\hat{a}$, with
possible outcomes $A\in\{+, -\}$. The expectation values of qubit 1 conditional
on $A$ are given (in the appendix) by
\begin{equation}\label{eq:conditional}
	\left.\vec\psi_1\right|_{A = \pm} = \frac{\vec\psi_1 \pm \Gamma\hat{a}}{1 \pm \vec\psi_2^T\hat{a}}\,,
\end{equation}
where we also note that conditionals (as opposed to marginals) of a pure
two-qubit state $\ket{\psi}$ are pure single-qubit states:
\begin{equation}\label{eq:conditional_normed}
	\abs{\left.\vec\psi_1\right|_{A=\pm}} = 1\,.
\end{equation}
We visualize this construction of conditionals in \cref{fig:1}C for the two
possible outcomes along one measurement axis; and in \cref{fig:1}D for a set of
potential measurement outcomes distributed over the whole qubit 2 Bloch sphere
in a spiral, illustrating the structure of the transformation for different
levels of entanglement.

\begin{figure}
	\includegraphics{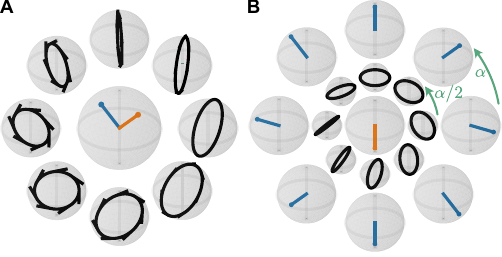}
	\caption{
Exploring parameter space.
(A) Freedom left in $\bar{R}$ if both marginals are fixed. For ease of
visualization, we plot the two (fixed) marginals in the same Bloch sphere in
the center and distinguish states only by the rotoreflection part.
(B) Rotating $\vec\psi_1$ (blue, outer ring) by rotating the mirror plane in
$\bar{R}$ (black, inner ring) through half the angle. As in (A), we plot
$\vec\psi_2$ just once in the center, since it remains fixed. This construction
might serve as intuition for the factor $1/2$ in the $SU(2)\to SO(3)$ mapping.
	}
	\label{fig:2}
\end{figure}

For a second application of our two-qubit picture, consider the freedom
remaining in a partially entangled state $\ket{\psi}$ upon fixing both
marginals (\cref{fig:2}A).
This remaining freedom must be associated with $\bar{R}$ alone (since
$\vec\psi_1$ and $\vec\psi_2$ are fixed by definition); it corresponds to the
$S^1$ fiber of \cref{thm:smooth} and acts as a rotation of the mirror plane,
with a simultaneous adjustment of the rotation angle in that plane such as to
maintain the fixed marginals. Note that the mirror plane returns to its initial
state already after half a rotation: $P_n = P_{-n} = \mathbb{1}_3 -
2\hat{n}\otimes\hat{n}$. This rotation of the mirror plane corresponds to the
local unitary transformation of rotating either qubit (a full turn!) around its
marginal.

A similar half-angle rotation effect is displayed in \cref{fig:2}B, where we
observe that---again due to the parity effect of $\bar{R}$---the Bloch vector
$\vec\psi_1$ can be rotated through an angle $\alpha$ by rotating the mirror
plane through only $\alpha/2$. This visualization might provide some intuition
for the factor $1/2$ in the usual mapping between $SU(2)$ (acting on states)
and $SO(3)$ (acting on expectation values / Bloch vectors).

Let our final observation about \cref{thm:bijective} be the following, by
direct calculation from \cref{tab:expectations}: the Bell basis states
correspond to $\bar{R} = P_x, P_y, P_z$ (the reflections along the coordinate
axes) for the triplet and $\bar{R} = P_0 = -\mathbb{1}_3$ for the singlet
states while of course $\vec\psi_1 = \vec\psi_2 = 0$ for all of them. The
depiction is rather trivial and thus omitted.

In summary, in this paper I propose a new visualization for pure two-qubit
states and prove that it behaves \enquote{nicely}, in that the underlying
parametrization is indeed diffeomorphic onto the space of observably different
states. The proposed picture provides intuition for entanglement in terms of
successive measurements of the qubits or the \enquote{hidden} degrees of
freedom when focusing only on the marginal Bloch vectors. Open questions not
addressed here include mixed states and three or more qubits, as well as an
intuitive understanding of the CNOT gate as the prototypical non-local unitary
transformation. Nonetheless, I hope that the presented picture can provide some
intuition for researchers and students alike.

\paragraph{Acknowledgements.} I am grateful to Aleksandra Galitsyna for
inspiring this study. R\'emy Mosseri, Stanislav Filatov, Avaneesh Narla, and
Eric Anschuetz provided valuable feedback.

\paragraph{Conflicts of interest.} The author declares no conflict of interest.

\paragraph{AI use.} I use generative AI to get a first impression of what
\enquote{is generally known} in a given field, as well as quickly falsifying
technical ideas through counterexamples. Beyond that, all the research,
logic \& calculations, writing, and graphics in this article were created
without AI; any potential errors are my own.

\appendix
\section{Appendix}

\subsection{Proof of \cref{thm:bijective}}

\begin{lemma}\label{thm:invertibility}
	The mapping $E: \mathcal{P}\to E(\mathcal{P})$ is invertible on its
	image. That is, from $\vec\psi_1$, $\vec\psi_2$, $\Gamma$ constructed
	as in \cref{thm:bijective}, we can uniquely identify $p\in\mathcal{P}$.

	\begin{proof}
		First note that $\vec w = \vec\psi_2$ can be directly read off
		from $E(p)$ and allows us to stratify: if $\abs{\vec w} < 1$,
		$p$ has to be in the $O^-(3)\times B^3$ component of
		$\mathcal{P}$, while $\abs{\vec w} = 1$ means $p\in S^2\times
		S^2$.

		Let thus $\abs{\vec w} < 1$. Then Sylvester's determinant
		theorem gives
		\begin{equation*}
			\det\left(\frac{\vec w\otimes\vec w}{1+\sqrt{1-\abs{\vec w}^2}} + \sqrt{1-\abs{\vec w}^2}\right) = 1-\abs{\vec w}^2 > 0\,,
		\end{equation*}
		so the matrix in brackets is invertible.  We can thus uniquely
		identify
		\begin{equation}
			\bar{R} = \Gamma \left(\frac{\vec w\otimes\vec w}{1+\sqrt{1-\abs{\vec w}^2}} + \sqrt{1-\abs{\vec w}^2}\right)^{-1}
		\end{equation}
		and assemble $p = (\bar{R}, \vec w) \in O^-(3)\times B^3$.

		Consider now $\abs{w} = 1$. We can directly read $\vec v =
		\vec\psi_1$ from $E(p)$ and thus uniquely identify $p = (\vec
		v, \vec w)\in S^2\times S^2$.

		We conclude that in either case, $E(p): \mathcal{P}\to
		E(\mathcal{P})$ is fully invertible on its image.
	\end{proof}
\end{lemma}

Note that we proved invertibility on the image of $E$, without making any
statement about how that image relates to other things we are interested in.
This is next.

Let
\begin{equation}
	\ket{\psi_0(\rho)} \equiv \cos\frac{\rho}{2}\ket{\uparrow\downarrow} + \sin\frac{\rho}{2}\ket{\downarrow\uparrow}\,.
\end{equation}
By inspection of \cref{tab:expectations} we find their expectation values
\begin{equation}
	\bra{\psi_0(\rho)}\sigma_\mu\otimes\sigma_\nu\ket{\psi_0(\rho)} \equiv \left[E^0(\rho)\right]_{\mu\nu}
\end{equation}
to be given by
\begin{equation}
	 E^0(\rho) = \begin{pmatrix} 1 & & & -\cos\rho \\ & \sin\rho & & \\ & & \sin\rho & \\ \cos\rho & & & -1 \end{pmatrix}\,.
\end{equation}
In terms of the decomposition of \cref{eq:decompE}, this reads
\begin{align*}
	\vec\psi_1^0 {}={}& \hat{z}\cos\rho \,, & \vec\psi_2^0 {}={}& -\hat{z}\cos\rho \,, & \Gamma^0 = P_zQ_z(\rho)\,,
\end{align*}
where $P_z \equiv\text{diag}\,(1, 1, -1) \in O^-(3)$ is the reflection along
the $z$-axis and 
\begin{equation*}
	Q_z(\rho) \equiv \sin\rho + \left(1-\sin\rho\right)\hat{z}\otimes\hat{z} = \text{diag}\,(\sin\rho, \sin\rho, 1)
\end{equation*}
a contraction in the $xy$-plane.

Note that $E^0(\rho)\in E(\mathcal{P})\,\forall\rho$. In fact, we can
generalize this statement to the following

\begin{lemma}\label{thm:rotations1}
	For any $\rho\in\left[0, \frac{\pi}{2}\right]$ and rotations $R_1,
	R_2\in SO(3)$, there exists a $p\in\mathcal{P}$ such that
	\begin{equation}
		\begin{pmatrix} 1 & \\ & R_1 \end{pmatrix} E^0(\rho) \begin{pmatrix} 1 & \\ & R_2^T \end{pmatrix} = E(p)\,.
	\end{equation}

	\begin{proof}
		Define the decomposition $\vec\psi_1$, $\vec\psi_2$, $\Gamma$
		for the matrix on the left hand side as in \cref{eq:decompE}
		and consider
		\begin{align}
			\vec\psi_1 {}={}& R_1\vec\psi_1^0 = R_1\hat{z}\cos\rho \equiv \vec v \,,\\
			\vec\psi_2 {}={}& R_2\vec\psi_2^0 = -R_2\hat{z}\cos\rho \equiv \vec w \,,\\
			\Gamma {}={}& R_1 \Gamma^0 R_2^T \\
			{}={}& R_1 P_z \left(\sin\rho + (1-\sin\rho) \hat{z}\otimes\hat{z} \right) R_2^T \\
			{}={}& R_1 P_z R_2^T \left(\sin\rho + \frac{1-\sin\rho}{\cos^2\rho}\vec\psi_2\otimes \vec\psi_2 \right) \\
			{}\equiv{}& \bar{R} \left(\frac{\vec w\otimes \vec w}{1 + \sqrt{1-\abs{\vec{w}}^2}} + \sqrt{1-\abs{\vec{w}}^2}\right)\,,
		\end{align}
		where we identified $\bar{R} \equiv R_1 P_z R_2^T$. Note that
		$\bar{R}\in O^-(3)$, since it is a product of $O(3)$ matrices
		with overall negative determinant. Furthermore certainly
		\begin{equation}
			\bar{R}\vec w = - R_1 P_z \hat{z}\cos\rho = R_1\hat{z}\cos\rho = \vec v\,,
		\end{equation}
		while for the case $\abs{\vec w} = 1$ the expression for
		$\Gamma$ indeed collapses to
		\begin{equation}
			\left.\Gamma\right|_{\abs{\vec w} = 1} = \bar{R}(\vec w \otimes \vec w) = \vec v \otimes \vec w\,.
		\end{equation}

		So in either case, we find one of the forms given for $E(p)$ in
		\cref{thm:bijective}: if $\rho = 0$ we have $p = (\vec v, \vec
		w)\in S^2\times S^2$ and accordingly $\Gamma = \vec v \otimes
		\vec w$; while for $\rho > 0$ we can set $\bar{R} \equiv R_1
		P_z R_2^T$ such that then $p = (\bar{R}, \vec w)\in
		O^-(3)\times B^3$ reproduces $\Gamma$ und $\vec\psi_1 =
		\bar{R}\vec \psi_2$. We have thus explicitly constructed a
		$p\in\mathcal{P}$ satisfying the \namecref{thm:rotations1}.
	\end{proof}
\end{lemma}

\Cref{thm:rotations1} gives us a means to construct matrices in
$E(\mathcal{P})$ by applying rotations to the expectation values of our anchor
states $\ket{\psi_0(\rho)}$. The converse also holds:

\begin{lemma}\label{thm:rotations}
	For any point $p\in\mathcal{P}$---that is: either $p\equiv(\bar{R},
	\vec w)\in O^-(3)\times B^3$, or $p\equiv(\vec v, \vec w)\in S^2\times
	S^2$---we can identify $\rho\in\left[0, \frac{\pi}{2}\right]$ and
	rotations $R_1, R_2\in SO(3)$ such that
	\begin{equation}
		\begin{pmatrix} 1 & \\ & R_1^T \end{pmatrix} E(p) \begin{pmatrix} 1 & \\ & R_2 \end{pmatrix} = E^0(\rho)\,.
	\end{equation}

	\begin{proof}
		Decompose $E(p)$ into $\vec\psi_1$, $\vec\psi_2$, $\Gamma$ as
		usual. By \cref{thm:invertibility}, from these we can also
		uniquely identify $p\in\mathcal{P}$; i.e., depending on
		$\abs{\vec w}$, $(\bar{R}, \vec w)\in O^-(3)\times B^3$ (in
		which case we also introduce the shorthand $\vec v =
		\bar{R}\vec w$) or $(\vec v, \vec w)\in S^2\times S^2$ (in
		which case $\bar{R}$ is not well-defined). Let
		\begin{equation}
			\rho \equiv \arccos\abs{\vec w}
		\end{equation}
		and decompose $\vec v$ and $\vec w$ into spherical coordinates:
		\begin{align}
			\vec v {}={}& R_z(\phi_1) R_y(\theta_1) \hat{z} \cos\rho\,, \\
			\vec w {}={}& -R_z(\phi_2) R_y(\theta_2) \hat{z} \cos\rho\,,
		\end{align}
		where we introduce the negative sign in the decomposition of
		$\vec w$ for convenience. Note that at this point we are
		somewhat indifferent about uniqueness of this decomposition; we
		only rely on its existence for any $\vec v, \vec w\in B^3$
		(including $0$, for which we can choose $\theta$, $\phi$
		arbitrary).

		This already resolves the branch $p\in S^2\times S^2$, i.e.\
		$\abs{\vec\psi_2} = 1$ (or $\rho = 0$). For in this case,
		setting
		\begin{equation}
			R_i \equiv R_z(\phi_i) R_y(\theta_i)
		\end{equation}
		gives
		\begin{align}
			R_1^T\vec\psi_1 {}={}& \hat{z} = \left.\vec\psi_1^0\right|_{\rho = 0} \,,\\
			R_2^T\vec\psi_2 {}={}& -\hat{z} = \left.\vec\psi_2^0\right|_{\rho = 0} \,,\\
			R_1^T\Gamma R_2 {}={}& R_1^T \left(\vec v \otimes \vec w\right) R_2 = -\hat{z}\otimes\hat{z} = \left.\Gamma^0\right|_{\rho = 0}\,.
		\end{align}

		For the rest of this proof, let thus $\abs{\vec\psi_2} < 1$,
		i.e.\ $p\in O^-(3)\times B^3$. So $\bar{R}$ is well-defined;
		consider thus
		\begin{equation}
			R \equiv \bar{R} R_z(\phi_2) R_y(\theta_2) P_z\,.
		\end{equation}
		By group multiplication in $O(3)$, we have $R\in O(3)$;
		furthermore, its determinant is
		\begin{equation}
			\det R = (-1)\cdot 1\cdot 1\cdot (-1) = 1
		\end{equation}
		so $R\in SO(3)$ is a rotation. We then have
		\begin{align}
			R_z(\phi_1) &R_y(\theta_1) \hat{z} \cos\rho = \vec v \equiv \bar{R}\vec w \\
			&{}={} -\bar{R} R_z(\phi_2) R_y(\theta_2) P_z^2 \hat{z}\cos\rho \\
			&{}={} -R P_z \hat{z}\cos\rho \\
			&{}={} R\hat{z}\cos\rho\,.
		\end{align}
		For $\abs{\vec\psi_2} > 0$ ($\rho < \frac{\pi}{2}$), this
		becomes
		\begin{equation}
			\left[ R_y^T(\theta_1) R_z^T(\phi_1) R \right] \hat{z} = \hat{z}\,,
		\end{equation}
		so the (trivially $SO(3)$ by group multiplication) matrix in
		square brackets is a rotation around an axis parallel to
		$\hat{z}$. It can thus be written as $R_z(\omega)$, giving
		\begin{equation}
			R = R_z(\phi_1) R_y(\theta_1) R_z(\omega)\,.
		\end{equation}
		Then setting
		\begin{align}
			R_1 {}={}& R_z(\phi_1) R_y(\theta_1) R_z(\omega) \equiv R \,, \\
			R_2 {}={}& R_z(\phi_2) R_y(\theta_2)
		\end{align}
		gives
		\begin{align}
			R_1^T\vec\psi_1 {}={}& R_z^T(\omega) \hat{z} \cos\rho = \vec\psi_1^0 \,,\\
			R_2^T\vec\psi_2 {}={}& -\hat{z} \cos\rho = \vec\psi_2^0 \,,\\
			R_1^T\Gamma R_2 {}={}& R^T \bar{R} \left(\frac{\vec w \otimes \vec w}{1 + \sin\rho} + \sin\rho\right) R_2 \\
			{}={}& R^T \bar{R} R_2 \left(\frac{\cos^2\rho}{1 + \sin\rho}\left(\hat{z} \otimes \hat{z}\right) + \sin\rho\right) \\
			{}={}& P_z Q_z(\rho) = \Gamma^0\,.
		\end{align}

		So finally, let $\abs{\vec\psi_2} = 0$, i.e.\ $\rho =
		\frac{\pi}{2}$. Let $R_1 = \bar{R} P_z \in SO(3)$ (as above:
		check the determinant) and $R_2 = \mathbb{1}$. Then
		\begin{align}
			R_1^T\vec\psi_1 {}={}& 0 = \left.\vec\psi_1^0\right|_{\rho = \frac{\pi}{2}} \,,\\
			R_2^T\vec\psi_2 {}={}& 0 = \left.\vec\psi_2^0\right|_{\rho = \frac{\pi}{2}} \,,\\
			R_1^T\Gamma R_2 {}={}& P_z \bar{R}^T \bar{R} = P_z = \left.\Gamma^0\right|_{\rho = \frac{\pi}{2}}\,.
		\end{align}

		For any point $p\in\mathcal{P}$ we have thus explicitly
		constructed $\rho$, $R_1$ and $R_2$ as in the \namecref{thm:rotations}.
	\end{proof}
\end{lemma}

Given that we know a state $\ket{\psi_0(\rho)}$ corresponding to the
expectations $E^0(\rho)$, can we use \cref{thm:rotations} to construct a state
$\ket{\psi}$ with expectations $E(p)$? Indeed, this is now almost trivial, by
observing that conjugation by single-qubit unitaries \enquote{rotates the Pauli
vector $\vec\sigma$}:
\begin{equation}
	\left[\ee^{\frac{\alpha}{2\ii}\hat{n}^T\vec\sigma}\right]^T \vec\sigma \left[\ee^{\frac{\alpha}{2\ii}\hat{n}^T\vec\sigma}\right] = R_n(\alpha)\vec\sigma\,,
\end{equation}
where $\hat{n}\in S^2$ is a unit vector, $\alpha\in\mathbb{R}$, and
$R_n(\alpha)\in SO(3)$ is the rotation by angle $\alpha$ around the axis
parallel to $\hat{n}$.  Multiplication on the left hand side of the equation
means $2\times 2$-matrix multiplication for each element of $\vec\sigma$; while
on the right hand side we take linear combinations of these elements as
described by a $3\times 3$-matrix $R_n(\alpha)$.

\begin{lemma}\label{thm:inE}
	$E(\mathcal{P}) \subset \mathcal{E}$. That is: for any
	$p\in\mathcal{P}$ there is some state $\ket{\psi}$ with expectations
	given by $E(p)$.

	\begin{proof}
		Let $p\in\mathcal{P}$ and identify $\rho$, $R_1$ and $R_2$ as
		in \cref{thm:rotations}. Write both in axis-angle notation:
		\begin{align}
			R_1 {}\equiv{}& R_m(\alpha) \,,& R_2 {}\equiv{}& R_n(\beta)
		\end{align}
		and set
		\begin{equation}
			\ket{\psi} \equiv \left[\ee^{\frac{\alpha}{2\ii}\hat{m}^T\vec\sigma} \otimes \ee^{\frac{\beta}{2\ii}\hat{n}^T\vec\sigma}\right] \ket{\psi_0(\rho)}\,.
		\end{equation}
		Then
		\begin{align}
			\E{\sigma_\mu \otimes \sigma_\nu} {}={}& \begin{pmatrix} 1 & \\ & R_m(\alpha) \end{pmatrix} \E{\sigma_\mu \otimes \sigma_\nu}_0 \begin{pmatrix} 1 & \\ & R_n^T(\beta) \end{pmatrix} \\
			{}={}& \begin{pmatrix} 1 & \\ & R_1 \end{pmatrix} E^0(\rho) \begin{pmatrix} 1 & \\ & R_2^T \end{pmatrix} \\
			{}={}& E(p)\,.
		\end{align}
	\end{proof}
\end{lemma}

\begin{lemma}\label{thm:any_state}
	$\mathcal{E}\subset E(\mathcal{P})$. That is: for any two qubit state
	$\ket{\psi}$ there is $p\in\mathcal{P}$ such that its expectations are
	given by $E(p)$.

	\begin{proof}
		Consider a generic state $\ket{\psi}$ and identify
		$\vec\psi_1$, $\vec\psi_2$, $\Gamma$ from its expectations as
		usual, as well as $\rho \equiv \arccos\abs{\vec\psi_2}$.
		Identify $R_m(\alpha), R_n(\beta)\in SO(3)$ such that
		\begin{align}
			\vec\psi_1 {}={}& R_m(\alpha)\hat{z}\cos\rho \,,\\
			\vec\psi_2 {}={}& -R_n(\beta)\hat{z}\cos\rho
		\end{align}
		and consider
		\begin{equation}
			\ket{\psi_R} \equiv \left[\ee^{-\frac{\alpha}{2\ii}\hat{m}^T\vec\sigma} \otimes \ee^{-\frac{\beta}{2\ii}\hat{n}^T\vec\sigma}\right] \ket{\psi}\,.
		\end{equation}
		Writing $E^{(R)}$ for the expectations of $\ket{\psi_{(R)}}$,
		we then have
		\begin{equation}\label{eq:ERE}
			E^R = \begin{pmatrix} 1 & \\ & R_m(-\alpha) \end{pmatrix} E \begin{pmatrix} 1 & \\ & R_n(-\beta) \end{pmatrix}\,.
		\end{equation}
		By construction, this gives $\vec\psi_1^R = \hat{z}\cos\rho$
		and $\vec\psi_2^R = -\hat{z}\cos\rho$.  Note that this
		construction is still valid for $\abs{\psi_2} = \cos\rho = 0$;
		in that case we are just free to choose the rotations as we
		wish, since the marginals are zero either way. So in that case
		we choose $R_m(\alpha) = \mathbb{1}_3 = R_n(\beta)$, such that
		$\ket{\psi_R} = \ket{\psi}$ and accordingly $E^R = E$.

		Now, write $\ket{\psi_R}$ in the computational basis as
		\begin{equation}
			\ket{\psi_R} = a\ket{\uparrow\uparrow} + b\ket{\uparrow\downarrow} + c\ket{\downarrow\uparrow} + d\ket{\downarrow\downarrow}
		\end{equation}
		and consult \cref{tab:expectations} for its expectation values.
		Since we already know that the marginals $\vec\psi_1$ and
		$\vec\psi_2$ need to point along $\pm\hat{z}$ (or be completely
		zero), this gives the following constraints on $a, b, c,
		d\in\mathbb{C}$:
		\begin{align}
			a^*b + c^*d {}={}& 0 \,,\label{eq:c1}\\
			a^*c + b^*d {}={}& 0 \,,\label{eq:c2}\\
			\abs{a}^2 + \abs{b}^2 - \abs{c}^2 - \abs{d}^2 {}={}& \cos\rho \,,\label{eq:c3}\\
			\abs{a}^2 - \abs{b}^2 + \abs{c}^2 - \abs{d}^2 {}={}& -\cos\rho \,.\label{eq:c4}
		\end{align}
		Adding $\labelcref{eq:c3} + \labelcref{eq:c4}$ gives $\abs{a} =
		\abs{d}$. Meanwhile, \cref{eq:c1} implies $\abs{ab} = \abs{cd}$
		which, together with $\abs{a} = \abs{d}$, becomes
		\begin{equation}\label{eq:cases}
			\Big(\abs{a} = \abs{d} = 0\Big) \lor \Big(\big(\abs{a} = \abs{d} \neq 0\big) \land \big(\abs{b} = \abs{c}\big)\Big)\,.
		\end{equation}

		We begin with the case $\abs{a} = \abs{d} = 0$. This
		trivializes \cref{eq:c1,eq:c2}, just leaving
		\begin{equation}
			\abs{b}^2 - \abs{c}^2 = \cos\rho\,;
		\end{equation}
		while state normalization gives $\abs{b}^2 + \abs{c}^2 = 1$.
		Putting both together, we get
		\begin{align}
			\abs{b} = \sqrt{\frac{1 + \cos\rho}{2}} = \cos\frac{\rho}{2} \,,\\
			\abs{c} = \sqrt{\frac{1 - \cos\rho}{2}} = \sin\frac{\rho}{2}
		\end{align}
		and thus there exist $\beta,\gamma\in S^1$ (not necessarily
		unique) such that
		\begin{align}\label{eq:bc}
			b {}={}& \ee^{\ii\beta}\cos\frac{\rho}{2}\,, & c {}={}& \ee^{\ii\gamma}\sin\frac{\rho}{2} \,.
		\end{align}
		The expectation values are insensitive to the global phase, so
		we introduce $\omega\equiv \gamma - \beta$. And indeed, using
		\cref{eq:bc} to read the expectation values $E^R$ from
		\cref{tab:expectations}, one finds
		\begin{align}
			E^R {}={}& \begin{pmatrix}
				1 & & & -\cos\rho \\
				  & \sin\rho\cos\omega & -\sin\rho\sin\omega & \\
				  & \sin\rho\sin\omega & \sin\rho\cos\omega & \\
				\cos\rho & & & -1
			\end{pmatrix} \\
				{}={}& \begin{pmatrix} 1 & \\ & R_z(\omega) \end{pmatrix}E^0(\rho)\,.
		\end{align}
		Together with \cref{eq:ERE}, this becomes
		\begin{equation*}
			E = \begin{pmatrix} 1 & \\ & R_m(\alpha) R_z(\omega) \end{pmatrix} E^0(\rho) \begin{pmatrix} 1 & \\ & R_n(\beta) \end{pmatrix} \equiv E(p)\,,
		\end{equation*}
		where the existence of $p\in\mathcal{P}$ is now guaranteed by
		\cref{thm:rotations1}.

		We thus continue with the second branch of \cref{eq:cases}
		which---by plugging back into \cref{eq:c3}---corresponds to the
		fully entangled states, $\rho = \frac{\pi}{2}$. In this case,
		our previous construction is essentially futile, since it was
		based on aligning $\psi_i^R = \pm\hat{z}\cos\rho$; which is
		trivially true for $\abs{\psi_i^R} = \cos\rho = 0$. The
		following is thus in fact an almost complete analysis of the
		fully entangled state set; the only cases missing are the ones
		with $\big(\abs{a} = \abs{d} = 0\big)\land\big(\abs{b} =
		\abs{c} = \frac{1}{\sqrt{2}}\big)$, which were quietly covered
		in the previous paragraph. Remember that above we set
		$\ket{\psi_R} = \ket{\psi}$ for this case, so we are dealing
		directly with our state of interest, without additional
		rotations as before.

		Let $\abs{a} = \abs{d} \equiv p \neq 0$ and $\abs{b} = \abs{c}
		\equiv q$. State normalization then gives $p^2 + q^2 =
		\frac{1}{2}$, which lets us define $\theta\in\left[0,
		\pi\right)$ such that
		\begin{align}
			\frac{1}{\sqrt{2}}\cos\frac{\theta}{2} {}\equiv{}& p = \abs{a} = \abs{d} \neq 0 \\
			\frac{1}{\sqrt{2}}\sin\frac{\theta}{2} {}\equiv{}& q = \abs{b} = \abs{c}\,.
		\end{align}
		We can then write the coefficients in polar coordinates as
		\begin{align}
			a {}={}& p\ee^{\ii\alpha}\,, & 
			b {}={}& q\ee^{\ii\beta}\,, & 
			c {}={}& q\ee^{\ii\gamma}\,, & 
			d {}={}& p\ee^{\ii\delta}\,. & 
		\end{align}
		For $\theta > 0$ (i.e.\ $q \neq 0 \neq p$), we then get an additional
		constraint on the phases from \cref{eq:c1,eq:c2}:
		\begin{equation}
			\ee^{\ii{\beta - \alpha}} + \ee^{\ii{\delta - \gamma}} = 0\,,
		\end{equation}
		which resolves as
		\begin{equation}
			\alpha - \beta - \gamma + \delta = \pm\pi\,.
		\end{equation}
		As before, the expectation values will be invariant under the
		global phase which, since $p\neq 0$, we can just choose to be
		$\alpha$. Let us thus introduce
		\begin{align}
			\phi_1 {}\equiv{}& \gamma - \alpha\,, & \phi_2 {}\equiv{}& \beta - \alpha\,,
		\end{align}
		giving accordingly
		\begin{equation}
			\delta - \alpha = \phi_1 + \phi_2 \pm \pi\,.
		\end{equation}
		We thus arrive at the full parametrization of states in this
		branch:
		\begin{align}
			a {}={}& \frac{\ee^{\ii\alpha}}{\sqrt{2}}\cos\frac{\theta}{2} \,, &
			b {}={}& \frac{\ee^{\ii(\alpha + \phi_2)}}{\sqrt{2}}\sin\frac{\theta}{2} \,, \label{eq:full1}\\
			c {}={}& \frac{\ee^{\ii(\alpha + \phi_1)}}{\sqrt{2}}\sin\frac{\theta}{2} \,, &
			d {}={}& -\frac{\ee^{\ii(\alpha + \phi_1 + \phi_2)}}{\sqrt{2}}\cos\frac{\theta}{2} \,. \label{eq:full2}
		\end{align}
		Note that the case $\theta = 0$ that was excluded above is also
		fully captured here, since $\alpha$ and $\alpha + \phi_1 +
		\phi_2$ are independent degrees of freedom (as usual, we are
		interested only in existence, not uniqueness of, in this case,
		$\phi_1$ and $\phi_2$). Likewise, the case $\big(\abs{a} =
		\abs{d} = 0\big) \land \big(\abs{b} = \abs{c} =
		\frac{1}{\sqrt{2}}\big)$ is covered by including $\pi$ in the
		domain of $\theta$, making \cref{eq:full1,eq:full2} a complete
		parametrization of the fully entangled states $\cos\rho = 0$;
		whose expectations can consequently be found by plugging
		\cref{eq:full1,eq:full2} back into \cref{tab:expectations},
		producing \cref{tab:expectations_R0}.
		\begin{table*}
		\setlength{\tabcolsep}{8pt}
		\setlength{\extrarowheight}{10pt}
		\caption{
			Expectation values for basis observables in fully
			entangled states $\rho = \frac{\pi}{2}$.
		}
		\label{tab:expectations_R0}
		\begin{tabular}{c||c c c c}
			$\E{\sigma_i\otimes\sigma_j}$ & $\mathbb{1}$ & $\sigma_x$ & $\sigma_y$ & $\sigma_z$ \\[5pt]
			\hline\hline
			$\mathbb{1}$ & 1 & 0 & 0 & 0 \\
			$\sigma_x$ & 0 & $\sin\phi_1\sin\phi_2 - \cos\theta\cos\phi_1\cos\phi_1$ & $-\sin\phi_1\cos\phi_2 - \cos\theta\cos\phi_1\sin\phi_2$ & $\sin\theta \cos\phi_1$ \\
			$\sigma_y$ & 0 & $-\cos\phi_1\sin\phi_2 - \cos\theta\sin\phi_1\cos\phi_2$ & $\cos\phi_1\cos\phi_2 - \cos\theta\sin\phi_1\sin\phi_2$ & $\sin\theta \sin\phi_1$ \\
			$\sigma_z$ & 0 & $\sin\theta \cos\phi_2$ & $\sin\theta \sin\phi_2$ & $\cos\theta$
		\end{tabular}
		\end{table*}
		By inspection, the nontrivial part of this matrix is a
		rotoreflection:
		\begin{equation}
			\Gamma = R_z(\phi_1) R_y(\theta) R_z(\phi_2) P_x \equiv \bar{R} \in O^-(3)\,.
		\end{equation}
		We can thus identify $p = (\bar{R}, 0) \in\mathcal{P}$ which
		reproduces all observables.

		In summary, for either branch, we have identified
		$p\in\mathcal{P}$ such that the expectations $E$ of
		$\ket{\psi}$ are given by $E(p)$.
	\end{proof}
\end{lemma}

\Cref{thm:invertibility,thm:inE,thm:any_state} together prove
\cref{thm:bijective}: $E(\mathcal{P})\subset \mathcal{E} \subset
E(\mathcal{P})$ implies $E(\mathcal{P}) = \mathcal{E}$; while invertibility on
the image then means that $E:\mathcal{P}\to\mathcal{E}$ is indeed bijective.
\qed

\subsection{Hopf fibration of $O^-(3)\times S^2$}

Given a reference unit vector $w\in S^2$, we can decompose any rotoreflection
$\bar{R}\in O^-(3)$ into a rotation around $w$ by some angle $\alpha\in S^1$,
followed by a reflection that maps $w$ to some new unit vector $v \equiv
\bar{R}w\in S^2$. This decomposition is invertible and \enquote{behaves
nicely}, in that it is a local diffeomorphism between $O^-(3)\times S^2$ and
$S^2\times S^2\times S^1$ (so it can be applied smoothly on open subsets, not
just in single points).

Intuitively (for readers with intuition in differential topology), these
statements follow by identifying the rotoreflections $O^-(3)$ with the
reflections $SO(3)$ (e.g. as $R \equiv -\bar{R}$), reflections with unit
quaternions modulo sign, applying the Hopf fibration, and resolving the sign
ambiguity in the fiber:
\begin{equation*}
	O^-(3) \to SO(3) \to S^3/\{\pm\} \to S^2\times \left(S^1/\{\pm\}\right) \to S^2\times S^1\,.
\end{equation*}

We give the precise statement in the following \namecref{thm:fibration}. Note
that for the proof we do not follow the sketched argument above too closely;
instead we just construct the bundle maps explicitly, since we will reuse them
in other proofs.

\begin{lemma}\label{thm:fibration}
	Let
	\begin{equation}
		\pi: O^-(3)\times S^2 \to S^2\times S^2: \pi(\bar{R}, w) = (\bar{R}w, w)\,.
	\end{equation}
	Then
	\begin{equation}
		S^1\longrightarrow O^-(3)\times S^2 \overset{\pi}{\longrightarrow} S^2\times S^2
	\end{equation}
	is a smooth fiber bundle.

	\begin{proof}
		Consider open subsets $V\subset S^2\cap\{z > 0\}$,
		$W\subset S^2\cap\{z < 0\}$ of the upper and lower (open)
		semisphere, respectively. Trivially, $V\cap W = \emptyset$ and
		$W\cap -W = \emptyset$. For $v\in V$, $w\in W$ let
		\begin{equation}\label{eq:Pvw}
			P(v, w) \equiv \mathbb{1}_3 - \frac{(v-w)\otimes(v-w)}{1-v^T w} \in O^-(3)
		\end{equation}
		be the reflection along the vector $v-w$ and note that specifically
		\begin{align}
			P(v, w)v {}={}& w\,, \\
			P(v, w)w {}={}& v\,, \\
			P^2(v, w) {}={}& \mathbb{1}_3\,.
		\end{align}
		$P(v, w)$ is well-defined iff $v\neq w$ (as guaranteed by
		$V\cap W = \emptyset$) and smooth where it is well-defined; so
		$P: V\times W\to O^-(3) \subset \mathbb{R}^{3\times 3}$ is a
		smooth map.

		The projection $\pi: (\bar{R}, w)\mapsto (\bar{R}w, w)$ of the
		\namecref{thm:fibration} is also smooth (composition of
		smooth maps). Let $p\equiv(\bar{R}, w)\in\pi^{-1}(V\times W)$,
		i.e.\ $\pi(p) = (\bar{R}w, w) \in V\times W = \text{dom}\,P$.
		We can then define (smoothly)
		\begin{equation}\label{eq:R_from_PRbar}
			R(\bar{R}, w) \equiv (P\circ\pi)(\bar{R}, w)\bar{R} = P(\bar{R}w, w)\bar{R}\in SO(3)\,,
		\end{equation}
		where $R\in O(3)$ because $P, \bar{R}\in O^-(3)\subset O(3)$
		and $\det R = \det P \det \bar{R} = 1$; so indeed $R\in SO(3)$
		is a rotation. Since furthermore
		\begin{equation}
			Rw = P(\bar{R}w, w)\bar{R}w = w,
		\end{equation}
		the axis of rotation is collinear with $w$; without loss of
		generality, we can thus write $R\equiv R_w(\alpha)$, a rotation
		about axis $w$ by some angle $\alpha\in S^1$.  Standard
		axis-angle decomposition gives $\alpha_w:
		R_w(\alpha)\mapsto\alpha$ uniquely and smoothly as
		\begin{equation}\label{eq:angle}
			\alpha_w(R) \equiv \arg\left(\text{tr}\,R - 1 -\ii\,\text{tr}\,X_wR\right)\,,
		\end{equation}
		$(X_w)_{jk}\equiv \sum_{i=1}^3 w_i\epsilon_{ijk}$ being the
		cross product matrix of $w$ ($X_w p = w\times p\,\forall
		p\in\mathbb{R}^3$). By construction $\alpha_w(R_w(\alpha)) =
		\alpha$ and all the following maps are smooth:
		\begin{align}
			R_w&: S^1\to SO(3) \\
			\alpha_w&: SO(3)\supset \text{im}\,R_w \to S^1 \\
			R_W&: W\times S^1 \to SO(3): (w, \alpha) \mapsto R_w(\alpha) \,.
		\end{align}
		We can thus assemble a smooth map
		\begin{align}
			\alpha_{VW}&: \pi^{-1}(V\times W) \to S^1 \\
			\alpha_{VW}&(\bar{R}, w)\equiv \alpha_w\left(P(\bar{R}w, w)\bar{R}\right)
		\end{align}
		into the $S^1$ fiber, allowing immediately to construct a
		smooth trivialization
		\begin{align}
			\varphi_{VW}&: \pi^{-1}(V\times W) \to V\times W\times S^1 \\
			\varphi_{VW}& {}\equiv{} \pi\times\alpha_{VW} \\
			\varphi_{VW}&(\bar{R}, w) = \left(\bar{R}w, w, \alpha_w\left(P(\bar{R}w, w)\bar{R}\right)\right)\,. \label{eq:phi_VW}
		\end{align}
		By construction, this map is compatible with the projection
		$\pi$.

		We now construct the inverse. Let $(v, w, \alpha)\in V\times
		W\times S^1$ and define
		\begin{align}
			\varphi^*_{VW}&: V\times W\times S^1 \to O^-(3)\times W \\
			\varphi^*_{VW}&(v, w, \alpha) \equiv \left(P(v, w)R_W(w, \alpha), w\right)\,.
		\end{align}
		as a candidate for $\varphi^{-1}_{VW}$.	Since both $P$ and
		$R_W$ are smooth, so is $\varphi^*_{VW}$, such that now we just
		need to check that
		\begin{align}
			&\left(\varphi_{VW}^*\circ\varphi_{VW}\right)(\bar{R}, w) \\
			&{}={} \varphi_{VW}^*\left(\bar{R}w, w, \alpha_w\left(P(\bar{R}w, w)\bar{R}\right)\right) \\
			&{}={} \left(P(\bar{R}w, w)R_w\left(\alpha_w\left(P(\bar{R}w, w)\bar{R}\right)\right), w\right) \\
			&{}={} \left(P^2(\bar{R}w, w)\bar{R}, w\right) \\
			&{}={} \left(\bar{R}, w\right)
		\end{align}
		and
		\begin{align}
			&\left(\varphi_{VW}\circ\varphi_{VW}^*\right)(v, w, \alpha) \\
			&{}={} \varphi_{VW}\left(P(v, w)R_W(w, \alpha), w\right) \\
			&{}={} \begin{pmatrix} P(v, w)R_w(\alpha)w \\ w \\ \alpha_w\left(P(P(v, w)R_w(\alpha)w, w)P(v, w)R_w(\alpha)\right) \end{pmatrix}^T \\
			&{}={} \left(v, w, \alpha_w\left(P^2(v, w)R_w(\alpha)\right)\right) \\
			&{}={} (v, w, \alpha)\,.
		\end{align}
		So indeed $\varphi^*_{VW} = \varphi^{-1}_{VW}$, and $\varphi_{VW}$ is surjective:
		\begin{equation}
			\text{im}\,\varphi_{VW} = \text{dom}\,\varphi^*_{VW} = V\times W\times S^1\,.
		\end{equation}
		This makes $\varphi_{VW}: \pi^{-1}(V\times W)\to V\times
		W\times S^1$ a smooth bijection with smooth inverse, i.e.\ a
		diffeomorphism.

		We can define similar maps on an open cover of $O^-(3)\times
		S^2$. Let us choose $V$ and $W$ maximal (i.e.\ equal to their
		(open) semispheres, instead of just subsets) and define
		$R_0\equiv\mathbb{1}_3$ and
		\begin{equation}
			R_i \equiv R_z\left(\frac{2\pi i}{3}\right) R_y\left(\arccos\left(-\frac{1}{3}\right)\right)
		\end{equation}
		for $i = 1, 2, 3$. These are chosen such that $R_i\hat{z}$ are
		the four corners of a tetrahedron, thus guaranteeing that
		\begin{equation}\label{eq:full_cover}
			\bigcup_{i=0}^3 R_iV = S^2 = \bigcup_{i=0}^3 R_iW\,.
		\end{equation}
		We now define 16 open domains
		\begin{equation}
			U_{ij} \equiv R_iV\times R_jW \subset S^2\times S^2
		\end{equation}
		for which, by construction,
		\begin{equation}
			\bigcup_{ij} U_{ij} = S^2\times S^2\,.
		\end{equation}
		We assemble the mapping on each domain by rotating to
		$\pi^{-1}(V\times W)$; applying $\varphi_{VW}$; and rotating
		the result back again. Let
		\begin{align}
			\mu_{ij}&: \pi^{-1}(U_{ij}) \to \pi^{-1}(V\times W) \\
			\mu_{ij}&(\bar{R}, w_r) \equiv \left(R_i^T\bar{R}R_j, R_j^Tw_r\right)\,, \\
			\mu_{ij}^{-1}&: \pi^{-1}(V\times W) \to \pi^{-1}(U_{ij}) \\
			\mu_{ij}^{-1}&(\bar{R}, w) \equiv \left(R_i\bar{R}R_j^T, R_jw\right)
		\end{align}
		and verify that
		\begin{equation*}
			(\mu_{ij}^{-1}\circ\mu_{ij})(\bar{R}, w) = \left(R_iR_i^T\bar{R}R_jR_j^T, R_jR_j^Tw\right) = (\bar{R}, w)\,.
		\end{equation*}
		$\mu_{ij}$ is thus bijective and smooth in both directions
		(multiplication by constant matrices), i.e.\ a diffeomorphism.
		Likewise,
		\begin{align}
			\nu_{ij}&: V\times W \to U_{ij}: (v, w) \mapsto (R_iv, R_jw)
		\end{align}
		is diffeomorphic by construction. Then
		\begin{align}
			\varphi_{ij}&: \pi^{-1}(U_{ij}) \to U_{ij}\times S^1 \\
			\varphi_{ij}& \equiv (\nu_{ij}\times\text{id})\circ\varphi_{VW}\circ\mu_{ij}
		\end{align}
		is a diffeomorphism by concatenation of diffeomorphisms. Since
		\begin{equation}
			(\nu_{ij}\circ\pi\circ\mu_{ij})(\bar{R}, w) = (\bar{R}w, w) = \pi(\bar{R}, w)
		\end{equation}
		we see that
		\begin{align}
			\varphi_{ij} {}={}& \nu_{ij}\circ\left(\pi\times\alpha_{VW}\right)\circ\mu_{ij} \\
			{}={}& \left(\nu_{ij}\circ\pi\circ\mu_{ij}\right) \times \left(\alpha_{VW}\circ\mu_{ij}\right) \\
			{}\equiv{}& \pi \times \alpha_{ij}\,;
		\end{align}
		is again compatible with $\pi$, with
		\begin{equation}
			\alpha_{ij}\equiv \alpha_{VW}\circ\mu_{ij}\,.
		\end{equation}

		Finally, the projection $\pi: O^-(3)\times S^2\to S^2\times
		S^2$ is surjective: for any $\left(v, w\right)\in S^2\times
		S^2$ set
		\begin{equation}
			\bar{R} = \begin{cases} P(v, w) & \text{if $v\neq w$} \\ \mathbb{1}_3 - 2n\otimes n & \text{if $v = w$}\end{cases} \in O^-(3) \,,
		\end{equation}
		$w\perp n \in S^2$ being any unit vector perpendicular to $w$.
		By construction $\bar{R}w = v$ in either case, such that
		\begin{equation}
			\left(\bar{R}, w\right) \in \pi^{-1}(v, w) \neq \emptyset\,.
		\end{equation}
		$\pi$ is thus surjective on $S^2\times S^2$.

		In summary, we have explicitly constructed an open cover
		$\{U_{ij}\}$ of $S^2\times S^2$ and associated diffeomorphisms
		$\varphi_{ij}: O^-(3)\times S^2\supset \pi^{-1}(U_{ij})\to
		U_{ij}\times S^1$ compatible with the smooth, surjective
		projection $\pi$. By definition, $O^-(3)\times S^2$ is thus a
		smooth $S^1$-fiber bundle over $S^2\times S^2$.
	\end{proof}
\end{lemma}

\begin{lemma}\label{thm:transitions}
	The transition maps between the local trivializations $\{(U_{ij},
	\varphi_{ij})\}$ of \cref{thm:fibration} are rigid rotations of the
	$S^1$ fiber:
	\begin{align}
		\varphi_{ij}\circ\varphi_{kl}^{-1}&: \left(U_{ij}\cap U_{kl}\right)\times S^1 \to \left(U_{ij}\cap U_{kl}\right)\times S^1: \\
		(v, w, \alpha)&{}\mapsto{} (v, w, \alpha + \Delta\alpha_{ijkl}(v, w))
	\end{align}
	with $\Delta\alpha_{ijkl}: \left(U_{ij}\cap U_{kl}\right)\to S^1$ smooth.

	\begin{proof}
		Smoothness of $\Delta\alpha_{ijkl}$ is trivial, since the
		$\varphi_{ij}$ are diffeomorphisms. We just need to show
		existence.

		Using the notation of (the proof of) \cref{thm:fibration}, consider
		\begin{align}
			\varphi_{ij}\circ\varphi_{kl}^{-1} {}={}& \left(\pi\circ\alpha_{ij}\right)\circ\varphi_{kl}^{-1} \\
			{}={}& \left(\pi\circ\varphi_{kl}^{-1}\right)\times\left(\alpha_{ij}\circ\varphi_{kl}^{-1}\right) \\
			{}={}& \text{proj}_{U_{kl}}\times \left(\alpha_{ij}\circ\varphi_{kl}^{-1}\right)\,,
		\end{align}
		so indeed the only non-trivial part of the transition map is in
		the fiber (as for any fiber bundle). Let thus $(v, w,
		\alpha)\in U_{ij}\cap U_{kl}\times S^1$ and assemble
		\begin{align*}
			\alpha'&{}\equiv{}\left(\alpha_{ij}\circ\varphi_{kl}^{-1}\right)\left(v, w, \alpha\right) \\
			&{}={} \left(\alpha_{VW}\circ\mu_{ij}\circ\mu_{kl}^{-1}\circ \varphi_{VW}^{-1}\circ \left(\nu_{kl}^{-1}\times\text{id}\right)\right)\left(v, w, \alpha\right) \\
			&{}={} \alpha_{w_j}\left(P(v_i, w_j)R_{ik}P(v_k, w_l)R_{w_l}(\alpha)R_{lj}\right) \\
			&{}\equiv{} \alpha_{w_j}(R_{jl}R_\Delta R_{w_l}(\alpha)R_{lj})\,,
		\end{align*}
		where
		\begin{align}
			(v_i, w_j) {}\equiv{}& \nu_{ij}^{-1}(v, w) = \left(R_i^Tv, R_j^Tw\right) \in V\times W \,,\\
			(v_k, w_l) {}\equiv{}& \nu_{kl}^{-1}(v, w) = \left(R_k^Tv, R_l^Tw\right) \in V\times W \,,\\
			R_{ik} {}\equiv{}& R_i^T R_k \,,\\
			R_{jl} {}\equiv{}& R_j^T R_l \,,\\
			R_\Delta {}\equiv{}& R_{lj}P(v_i, w_j)R_{ik}P(v_k, w_l)
		\end{align}
		and we use the shorthands $\alpha'\equiv \alpha'(v, w,
		\alpha)$, $R_\Delta \equiv R^\Delta_{ijkl}(v, w)$ for
		convenience. Note that $R_\Delta \in SO(3)$ by group
		multiplication in $O(3)$ and
		\begin{equation}
			\det R_\Delta = 1\cdot(-1)\cdot 1\cdot (-1) = 1\,.
		\end{equation}
		Since furthermore $R_\Delta w_l = w_l$ by construction of the
		factors, $R_\Delta\equiv R_{w_l}(\Delta\alpha)$ is in fact a
		rotation around $w_l$, which we can combine with the existing
		$R_{w_l}(\alpha)$ to give
		\begin{align}
			\alpha' {}={}& \alpha_{w_j}\left(R_{jl}R_{w_l}(\alpha+\Delta\alpha)R_{lj}\right) \\
			{}={}& \alpha_{w_j}\left(R_{w_j}(\alpha+\Delta\alpha)\right) \\
			{}={}& \alpha + \Delta\alpha\,.
		\end{align}
		Restoring full notation gives
		\begin{equation}
			\alpha'(v, w, \alpha) = \alpha + \Delta\alpha_{ijkl}(v, w)\,.
		\end{equation}
	\end{proof}
\end{lemma}

\subsection{Proof of \cref{thm:smooth}}\label{sec:proof:smooth}

Let $Q\equiv O^-(3)\times\widebar{B^3}$ be the domain of the quotient map $q$,
note that indeed
\begin{align}
	\text{im}\,q {}\equiv{}& q(Q) \\
	{}={}& O^-(3)\times B^3\sqcup S^2\times S^2 \equiv \mathcal{P} \\
	{}\simeq{}& O^-(3)\sqcup O^-(3)\times S^2\times (0, 1)\sqcup S^2\times S^2 \\
	{}\equiv{}& \mathcal{P}_0\sqcup\mathcal{P}_{(0, 1)} \sqcup\mathcal{P}_1\,,
\end{align}
and let likewise
\begin{align}
	\mathcal{P}_{[0, 1)} {}\equiv{}& \mathcal{P}_0\sqcup\mathcal{P}_{(0, 1)} = \mathcal{P}\setminus\mathcal{P}_1\,, \\
	\mathcal{P}_{(0, 1]} {}\equiv{}& \mathcal{P}_{(0, 1)}\sqcup\mathcal{P}_1 = \mathcal{P}\setminus\mathcal{P}_0\,.
\end{align}

The quotient map $q$ acts non-trivially only on $\partial Q = O^-(3)\times
S^2$, where it coincides with the bundle projection
\begin{equation}
	\pi: O^-(3)\times S^2\to S^2\times S^2: (\bar{R}, w)\mapsto (\bar{R}w, w)
\end{equation}
of the smooth fiber bundle
\begin{equation}
	S^1\longrightarrow O^-(3)\times S^2\overset{\pi}{\longrightarrow} S^2\times S^2
\end{equation}
constructed in \cref{thm:fibration}; we might write
\begin{equation}\label{eq:q_pi}
	q\circ\pi^{-1} = \text{id}_{S^2\times S^2}\,.
\end{equation}
This will allow us to smoothly extend that bundle to
\begin{equation}
	B^2\longrightarrow \mathcal{P}_{(0, 1]}\overset{\pi^*}{\longrightarrow} S^2\times S^2\,,
\end{equation}
as we show below; thus endowing $\mathcal{P}_{(0, 1]}$ with a smooth structure.

In the following, we use notation and constructions from the proof of
\cref{thm:fibration} without further comment; specifically: the local
trivializations $\{(U_{ij}, \varphi_{ij})\}$ and their inverses, the parity
operation $P(v, w)$, rotations $R_w(\alpha)$, and the corresponding inverse
$\alpha_w:\text{im}\,R_w\to S^1$ extracting the rotation angle.

We begin by constructing $\mathcal{P}_{(0, 1]}$ as a fiber bundle and return to
the question of smoothness afterwards.  Let
\begin{equation}
	\pi^*: \mathcal{P}_{(0, 1]}\to S^2\times S^2: \begin{cases} (\bar{R}, w, r)\mapsto (\bar{R}w, w) \\ (v, w)\mapsto (v, w) \end{cases}
\end{equation}
be the bundle projection, with the two branches corresponding to the two
disjoint components $\mathcal{P}_{(0, 1)}$ and $\mathcal{P}_1$, respectively.
$\pi^*$ is surjective, since $\pi^*(\mathcal{P}_1) = \mathcal{P}_1 = S^2\times
S^2$. Let $\mathcal{O}\subset S^2\times S^2$ be open and note that
\begin{equation}
	\left(\pi^*\right)^{-1}(\mathcal{O}) = \pi^{-1}(\mathcal{O})\times (0, 1)\sqcup \mathcal{O} \label{eq:tmp1}\,.
\end{equation}
Then
\begin{align}
	\left[q^{-1}\circ\left(\pi^*\right)^{-1}\right](\mathcal{O}) {}={}& \pi^{-1}(\mathcal{O})\times (0, 1)\sqcup q^{-1}(\mathcal{O}) \\
	{}\simeq{}& \pi^{-1}(\mathcal{O})\times (0, 1] \label{eq:dom1}
\end{align}
is open in $Q$ by continuity of $\pi$; in the last step, we exploited
\cref{eq:q_pi} to write $q^{-1}(\mathcal{O}) = \pi^{-1}(\mathcal{O})$. By
definition of the quotient topology on $\mathcal{P}$, $\pi^*$ is thus
continuous.

Let
\begin{align}
	z_{ij}^*&: \left(\pi^*\right)^{-1}(U_{ij})\to B^2 \subset \mathbb{C} \\
	&\begin{cases}
		(\bar{R}, w, r)\mapsto \sqrt{1-r^2}\,\ee^{\ii\alpha_{ij}(\bar{R}, w)} \\
		(v, w)\mapsto 0
	\end{cases}\,.
\end{align}
From \cref{eq:tmp1} and $\alpha_{ij}\left(\pi^{-1}(U_{ij})\right) = S^1$, we
conclude that $z_{ij}^*$ is surjective. It is also continuous: let
$\mathcal{O}\subset B^2$ be open. If $0\not\in\mathcal{O}$,
$\left(z_{ij}^*\right)^{-1}(\mathcal{O})\subset\mathcal{P}_{(0, 1)}$ is open by
continuity of $\alpha_{ij}$ and elementary functions. Let thus without loss of
generality $0\in\mathcal{O} = B^2_\epsilon$, the open unit disk of radius
$\epsilon\in (0, 1)$. Then again,
\begin{align}
	&\left[q^{-1} \circ \left(z_{ij}^*\right)^{-1}\right]\left(B^2_\epsilon\right) \\
	&{}={} q^{-1}\left(\pi^{-1}(U_{ij})\times \left(\sqrt{1-\epsilon^2}, 1\right)\sqcup U_{ij}\right) \\
	&{}={} \pi^{-1}(U_{ij})\times \left(\sqrt{1-\epsilon^2}, 1\right) \sqcup \pi^{-1}(U_{ij}) \\
	&{}\simeq{} \pi^{-1}(U_{ij})\times \left(\sqrt{1-\epsilon^2}, 1\right]
\end{align}
is open in $Q$ by continuity of $\pi$.  Thus, by definition, $z_{ij}^*$ is
continuous.

Let now
\begin{align}
	\varphi_{ij}^*&: \left(\pi^*\right)^{-1}(U_{ij})\to U_{ij}\times B^2 \\
	\varphi_{ij}^*& \equiv \pi^*\times z_{ij}^* \\
	&\begin{cases}
		(\bar{R}, w, r)\mapsto \left(\bar{R}w, w, \sqrt{1-r^2}\,\ee^{\ii\alpha_{ij}}\right) \\
		(v, w)\mapsto (v, w, 0)
	\end{cases}\,,
\end{align}
surjective, continuous, and compatible with $\pi^*$ by construction. The
inverse is easily identified as
\begin{equation}
	\left(\varphi_{ij}^*\right)^{-1}\left(v, w, 0\right) \equiv (v, w)
\end{equation}
and
\begin{equation}
	\left(\varphi_{ij}^*\right)^{-1}\left(v, w, z\right) = \left(\varphi_{ij}^{-1}(v, w, \arg z), \sqrt{1-\abs{z}^2}\right)
\end{equation}
if $z\neq 0$. Regarding continuity of $\left(\varphi_{ij}^*\right)^{-1}$, note
that
\begin{align*}
	\varphi_{ij}^*\circ q:{} &\pi^{-1}(U_{ij})\times (0, 1]\to U_{ij}\times B^2: \\
	(\bar{R}, w, r){}\mapsto{}& \begin{cases}
		\left(\bar{R}w, w, \sqrt{1-r^2}\,\ee^{\ii\alpha_{ij}(\bar{R}, w)}\right) & \text{if $r < 1$} \\
		(\bar{R}w, w, 0) & \text{if $r = 1$}
	\end{cases} \\
	&{}={} \left(\pi(\bar{R}, w), \sqrt{1-r^2}\,\ee^{\ii\alpha_{ij}(\bar{R}, w)}\right)\,,
\end{align*}
where we used \cref{eq:dom1} to identify the domain. Over $\pi^{-1}(U_{ij})$,
the bundle projection $\pi$ acts as the trivial projection $U_{ij}\times S^1\to
U_{ij}$, which is an open map; we thus only need to check the $B^2$ component
in detail.  Consider thus
$\mathcal{O}\subset\left(\pi^*\right)^{-1}(U_{ij})\subset\mathcal{P}_{(0, 1]}$
open, i.e.\ $q^{-1}(\mathcal{O})\subset Q$ open, and note that
\begin{equation}
	q^{-1}(\mathcal{O}) \supset q^{-1}\left(\mathcal{O}\cap\mathcal{P}_1\right) = \pi^{-1}\left(\mathcal{O}\cap\mathcal{P}_1\right)\,.
\end{equation}
That is, for any point $p\in\mathcal{O}\cap\mathcal{P}_1$ coming from the
$\mathcal{P}_1$ component, $q^{-1}(\mathcal{O})$ contains the whole $S^1$-fiber
over $p$ in $\partial Q$. Since $q^{-1}(\mathcal{O})$ is open, it then also
contains an open cover of the fiber; which, because the fiber is compact, has a
finite subcover and can thus be written as
\begin{equation}
	\pi^{-1}(p)\times\{1\} \subset U_p\times (\epsilon, 1]\subset q^{-1}(\mathcal{O})
\end{equation}
without loss of generality, for some $\epsilon > 0$ and $U_p\subset
O^-(3)\times S^2$ open. We can thus check that the pre-image
\begin{align}
	\left[\left(\varphi_{ij}^*\right)^{-1}\right]^{-1}(\mathcal{O}) {}={}& \left(\varphi_{ij}^*\circ q\right)\left(q^{-1}(\mathcal{O})\right) \\
	{}\supset{}& \left(\varphi_{ij}^*\circ q\right)\left(U_p\times(\epsilon, 1]\right) \\
	{}={}& \pi(U_p)\times B^2_{\sqrt{1-\epsilon^2}}\,,
\end{align}
contains an open neighborhood of $\varphi_{ij}^*(p) \equiv (p, 0)$, so
$\left(\varphi_{ij}^*\right)^{-1}$ is continuous in $(p, 0)\in U_{ij}\times
B^2$. Since $p\in\mathcal{O}\cap\mathcal{P}_1$ and
$\mathcal{O}\subset\left(\pi^*\right)^{-1}(U_{ij})$ itself were arbitrary,
$\left(\varphi_{ij}^*\right)^{-1}$ is continuous everywhere on $U_{ij}\times
\{0\}$. With continuity in $(v, w, z)$ for $z\neq 0$ trivial by continuity of
$\varphi_{ij}^{-1}$, $\arg:\mathbb{C}\setminus\{0\}\to S^1$, and elementary
functions, we conclude that $\left(\varphi_{ij}^*\right)^{-1}$ is continuous
everywhere. This makes the $\varphi_{ij}^*$ homeomorphisms and
\begin{equation}
	B^2\longrightarrow\mathcal{P}_{(0, 1]}\overset{\pi^*}{\longrightarrow} S^2\times S^2
\end{equation}
a fiber bundle.

We now revisit the smooth structure of $\mathcal{P}_{(0, 1]}$.  Since both the
base space $S^2\times S^2$ and fiber $B^2$ are smooth submanifolds of
$\mathbb{R}^n$ ($n = 6, 2$, respectively), we can construct a smooth atlas on
$\mathcal{P}_{(0, 1]}$ from the local trivializations, iff the transition maps
between them are smooth. To that end, consider
\begin{align}
	&\left[z_{ij}^*\circ \left(\varphi_{kl}^*\right)^{-1}\right](v, w, z) \\
	&{}={} \begin{cases}
		\abs{z}\exp\left(\ii\left[\alpha_{ij}\circ\varphi_{kl}^{-1}\right] (v, w, \arg z)\right) & \text{if $z\neq 0$} \\
		0 & \text{if $z = 0$}
	\end{cases} \\
	&{}={} z\exp\left(\ii\Delta \alpha_{ijkl}\left(v, w\right)\right) \,,
\end{align}
where
\begin{equation}
	\left[\alpha_{ij}\circ\varphi_{kl}^{-1}\right](v, w, \arg z) = \arg z + \Delta\alpha_{ijkl}(v, w)
\end{equation}
by \cref{thm:transitions}. Thus the transition maps
\begin{equation}
	\varphi_{ij}^*\circ\left(\varphi_{kl}^*\right)^{-1} = \text{proj}_{U_{kl}}\times\left[z_{ij}^*\circ\left(\varphi_{kl}^*\right)^{-1}\right]
\end{equation}
are smooth, endowing $\mathcal{P}_{(0, 1]}$ with the structure of a smooth
fiber bundle.

At the other end of the radial coordinate, $\mathcal{P}_{[0, 1)} = O^-(3)\times
B^3$ is naturally smooth as a product of smooth manifolds. Its smooth structure
is compatible with the bundle structure on $\mathcal{P}_{(0, 1]}$ by
decomposing $B^3\setminus\{0\}$ into spherical coordinates:
\begin{equation}\label{eq:P01_identification}
	O^-(3)\times B^3\setminus\{0\}\simeq \mathcal{P}_{(0, 1)}\simeq O^-(3)\times S^2\times (0, 1)\,,
\end{equation}
which is diffeomorphic on $\mathcal{P}_{(0, 1)} = \mathcal{P}_{[0, 1)} \cap
\mathcal{P}_{(0, 1]}$. We can thus smoothly glue these two manifolds on their
overlap, giving
\begin{equation}
	\mathcal{P}_{[0, 1)}\cup\mathcal{P}_{(0, 1]} = \mathcal{P}
\end{equation}
the structure of a smooth manifold. Since neither of the components has a
boundary, $\partial\mathcal{P} = \emptyset$.  \qed

\subsection{Proof of \cref{thm:diffeo}}

With the work done up until here, this becomes relatively straightforward.
Since we already know from \cref{thm:bijective} that
$E:\mathcal{P}\to\mathcal{E}$ is bijective, we only need to show smoothness and
smoothness of the inverse. We decompose
\begin{equation}
	\mathcal{P} = \mathcal{P}_{[0, 1)} \cup \mathcal{P}_{(0, 1]}
\end{equation}
as in (the proof of) \cref{thm:smooth} and treat the cases separately,
exploiting the smooth structure on each:
\begin{equation}
	\mathcal{P}_{[0, 1)} = O^-(3)\times B^3
\end{equation}
and
\begin{equation}
	B^2 \longrightarrow \mathcal{P}_{(0, 1]}\overset{\pi^*}{\longrightarrow} S^2\times S^2\,,
\end{equation}
i.e.\ $\mathcal{P}_{(0, 1]} \simeq S^2\times S^2\times B^2$ locally.

Let thus $U_{ij}$, $z^*_{ij}: \left(\pi^*\right)^{-1}(U_{ij})\to B^2$, and
$\varphi_{ij}^* = \pi^*\times z_{ij}^*$ as in (the proof of) \cref{thm:smooth}
above. Take $\left(\hat{v}_r, \hat{w}_r, z\right)\in U_{ij}\times B^2$ and
write $\hat{v} \equiv R_i^T\hat{v}_r$, $\hat{w} \equiv R_j^T\hat{w}$. Consider
then
\begin{align*}
	&\left(\varphi_{ij}^*\right)^{-1}(\hat{v}_r, \hat{w}_r, z) \\
	&{}={} \begin{cases}
		\left(\varphi_{ij}^{-1}(\hat{v}_r, \hat{w}_r, \arg z), \sqrt{1-\abs{z}^2}\right) & \text{if $\abs{z} > 0$} \\
		\left(\hat{v}_r, \hat{w}_r\right) & \text{if $\abs{z} = 0$}
	\end{cases} \\
	&{}={} \begin{cases}
		\left(R_iP(\hat{v}, \hat{w})R_w(\arg z)R_j^T, R_j\hat{w}, \sqrt{1-\abs{z}^2}\right) \\
		\left(R_i\hat{v}, R_j\hat{w}\right)
	\end{cases} \\
	&{}\simeq{} \begin{cases}
		\left(R_iP(\vec{v}, \vec{w})R_w(\arg z)R_j^T, R_j\vec{w}\right) \\
		\left(R_i\vec{v}, R_j\vec{w}\right)
	\end{cases}\,,
\end{align*}
where in the last step we use the identification \cref{eq:P01_identification}
to introduce
\begin{align}
	\vec v {}\equiv{}& \sqrt{1-\abs{z}^2}\,\hat{v} \,, &
	\vec w {}\equiv{}& \sqrt{1-\abs{z}^2}\,\hat{w}
\end{align}
and omit the branch conditions after the first split for convenience. We assemble
\begin{equation}
	E\circ\left(\varphi_{ij}^*\right)^{-1}: U_{ij}\times B^2 \to \mathcal{E}
\end{equation}
in the usual decomposition $\vec\psi_1$, $\vec\psi_2$, and $\Gamma$ directly
from \cref{thm:bijective}:
\begin{widetext}
\begin{align}
	\vec\psi_1(\hat{v}_r, \hat{w}_r, z) {}\equiv{}& R_i \vec v = \sqrt{1-\abs{z}^2}\,\hat{v}_r \,,\\
	\vec\psi_2(\hat{v}_r, \hat{w}_r, z) {}\equiv{}& R_j \vec w = \sqrt{1-\abs{z}^2}\,\hat{w}_r \,,\\
	\Gamma(\hat{v}_r, \hat{w}_r, z) {}={}& \begin{cases}
		R_i P(\vec v, \vec w)R_w(\arg z) R_j^T \left(\dfrac{(R_j\vec w)\otimes (R_j\vec w)}{1+\sqrt{1-\abs{R_j\vec w}}^2} + \sqrt{1-\abs{R_j\vec{w}}^2} \right) \\
		(R_i\vec v) \otimes (R_j\vec w)
	\end{cases} \\
	{}={}& R_i P(\vec v, \vec w) \left\{\begin{aligned}
		&\dfrac{\vec w\otimes \vec w}{1+\abs{z}} + \abs{z}R_w(\arg z) & \text{(if $\abs{z} > 0$)}\\
		&\vec w\otimes \vec w & \text{(if $\abs{z} = 0$)}
	\end{aligned}\right\} R_j^T \\
	{}={}& R_i P(\vec v, \vec w) \left\{\begin{aligned}
		&(1 - \abs{z})\hat{w}\otimes\hat{w} + \abs{z}\hat{w}\otimes\hat{w} + \abs{z}\left(1 - \hat{w}\otimes\hat{w}\right)\cos\arg z + \abs{z}X_w\sin\arg z \\
		& \hat{w}\otimes\hat{w}
	\end{aligned}\right\} R_j^T \label{eq:rodriguez} \\
	{}={}& R_i P(\vec v, \vec w) \left[ \hat{w}\otimes\hat{w} + (1-\hat{w}\otimes\hat{w})\Re z + X_w\Im z \right] R_j^T \label{eq:Gnice} \,.
\end{align}
\end{widetext}
In step \labelcref{eq:rodriguez} we use Rodriguez' formula for axis-angle
rotations:
\begin{equation}
	R_n(\alpha) = \hat{n}\otimes \hat{n} + (1 - \hat{n}\otimes \hat{n})\cos\alpha + X_n\sin\alpha\,,
\end{equation}
where $\hat{n}\in S^2$, $\alpha\in S^1$, and $X_n$ is the cross product matrix
of $\hat{n}$ as in \cref{eq:angle}, whose components are given by
$\left(X_n\right)_{ac} = \epsilon_{abc}w^b$.

With $\vec w$, $\hat{w}$, and $X_w$ linear (and thus smooth) in $\hat{w}$;
$\vec v$ and $\hat{v}$ likewise linear (and thus smooth) in $\hat{v}$; and
$\vec v$, $\vec w$ smooth in $(\Re z, \Im z)\in B^2\subset \mathbb{R}^2$ (not
holomorphic in $z\in B^2\subset\mathbb{C})$, we find that $\vec\psi_1$,
$\vec\psi_2$, and $\Gamma$ are elementary combinations of smooth functions and
thus smooth over $U_{ij}\times B^2$. Since this holds for all the
trivializations $\varphi_{ij}^*$ of the smooth fiber bundle $\mathcal{P}_{(0,
1]}$, the mapping $E: \mathcal{P}_{(0, 1]}\to\mathcal{E}$ is smooth.

Construct thus the inverse, $\varphi_{ij}^* \circ E^{-1}$. First, note that
\begin{align}
	\hat{v}_r(\vec\psi_1, \vec\psi_2, \Gamma) {}={}& \frac{\vec\psi_1}{\abs{\vec\psi_1}}\,, \\
	\hat{w}_r(\vec\psi_2, \vec\psi_2, \Gamma) {}={}& \frac{\vec\psi_2}{\abs{\vec\psi_2}}
\end{align}
are smooth, since $\vec\psi_i \neq 0$ over $E(\mathcal{P}_{(0, 1]})$. Thus
likewise
\begin{align*}
	\hat{v} {}\equiv{}& \frac{R_i^T\vec\psi_1}{\abs{\vec\psi_1}} \,,& 
	\vec{v} {}\equiv{}& R_i^T\vec\psi_1 \,,&
	\hat{w} {}\equiv{}& \frac{R_j^T\vec\psi_2}{\abs{\vec\psi_2}} \,,& 
	\vec{w} {}\equiv{}& R_j^T\vec\psi_2
\end{align*}
are smooth. We therefore need to check $z(\vec\psi_1, \vec\psi_2, \Gamma)$,
which we extract componentwise from $\Gamma$. Consider, to that end,
\begin{align}
	\text{tr} \left[P(\vec v, \vec w) R_i^T \Gamma R_j \right] {}={}& 1 + 2\Re z \,,\\
	\text{tr} \left[X_w P(\vec v, \vec w) R_i^T \Gamma R_j \right] {}={}& -2\Im z \,,
\end{align}
where we used $\text{tr}\, X_w = 0$ and $X_w^2 = \hat{w}\otimes\hat{w} - 1$. We
can thus extract
\begin{equation*}
	z(\vec\psi_1, \vec\psi_2, \Gamma) = \frac{1}{2}\left(\text{tr}\left[\left(\mathbb{1}_3 - \ii X_w\right) P(\vec v, \vec w) R_i^T \Gamma R_j\right] - 1\right)\,,
\end{equation*}
which is smooth in $\vec\psi_1$, $\vec\psi_2$ and $\Gamma$ over
$E(\mathcal{P}_{(0, 1]})$. Over $\mathcal{P}_{(0, 1]}$, we have thus shown that
both $E$ and its inverse are smooth.

Consider now the second part, $E:\mathcal{P}_{[0, 1)}\to E(\mathcal{P}_{[0,
1)})$. With $\mathcal{P}_{[0, 1)} \equiv O^-(3)\times B^3$ carrying the natural
smooth structure inherited from $\mathbb{R}^{3\times 3}\times\mathbb{R}^3
\simeq \mathbb{R}^{12}$, this is quite straightforward: \cref{thm:bijective}
gives (on this restricted domain)
\begin{align}
	\vec\psi_1(\bar{R}, \vec w) {}={}& \bar{R}\vec w\,, \\
	\vec\psi_2(\bar{R}, \vec w) {}={}& \vec w\,, \\
	\Gamma(\bar{R}, \vec w) {}={}& \bar{R}\left(\frac{\vec w\otimes\vec w}{1 + \sqrt{1-\abs{\vec w}^2}} + \sqrt{1-\abs{\vec w}^2} \right)\,.
\end{align}
Since the critical point $\abs{w} = 1$ of the square root in $\Gamma$ is not in
the domain, this is trivially smooth. Likewise, \cref{thm:invertibility} gives
\begin{align*}
	\bar{R}(\vec\psi_1, \vec\psi_2, \Gamma) {}={}& \Gamma\left(\frac{\vec\psi_2\otimes\vec\psi_2}{1 + \sqrt{1-\abs{\vec\psi_2}^2}} + \sqrt{1-\abs{\vec\psi_2}^2} \right)^{-1}\,, \\
	\vec w(\vec\psi_1, \vec\psi_2, \Gamma) {}={}& \vec\psi_2\,,
\end{align*}
where invertibility of the matrix in brackets is again guaranteed by
$\abs{\vec\psi_2} < 1$. This is also a concatenation of smooth elementary
functions and thus smooth.

In summary, we have shown that $E:\mathcal{P}\to \mathcal{E}$ acts smoothly and
with a smooth inverse on both components $\mathcal{P}_{[0, 1)}$ and
$\mathcal{P}_{(0, 1]}$, which glue smoothly by \cref{thm:smooth}; by
\cref{thm:bijective} it is also bijective. We thus conclude that
$E:\mathcal{P}\to\mathcal{E}$ is a diffeomorphism. \qed

\subsection{Derivation of \cref{eq:conditional,eq:conditional_normed}}

From a purely statistical point of view, a state $\ket{\psi}$ together with two
measurement axes $\hat{a}$ (for qubit 2) and $\hat{b}$ (for qubit 1) defines a
joint probability distribution $p(A, B)$ over the possible measurement outcomes
$\left(A, B\right)\in\{\pm\}^2$. We have
\begin{align*}
	1 {}={}& p(+, +) + p(+, -) + p(-, +) + p(-, -) \,,\\
	\E{A} {}={}& \E{\mathbb{1}_2 \otimes \left(\hat{a}^T\vec\sigma\right)} \equiv \vec\psi_2^T\hat{a} \\
	{}={}& p(+, +) + p(+, -) - p(-, +) - p(-, -)\,,\\
	\E{B} {}={}& \E{\left(\hat{b}^T\vec\sigma\right) \otimes \mathbb{1}_2} \equiv \hat{b}^T\vec\psi_1 \\
	{}={}& p(+, +) - p(+, -) + p(-, +) - p(-, -)\,,\\
	\E{AB} {}={}& \E{\left(\hat{b}^T\vec\sigma\right) \otimes \left(\hat{a}^T\vec\sigma\right)} \equiv \hat{b}^T \Gamma \hat{a} \\
	{}={}& p(+, +) - p(+, -) - p(-, +) + p(-, -)\,,
\end{align*}
providing a connection between the statistics in terms of $p(A, B)$ and the
physics in terms of $\ket{\psi}$ and $\vec\sigma$. Naturally, we can invert
this system to identify $p(A, B)$ in terms of the expectation values and thus
$\ket{\psi}$; letting us directly write the conditional expectations
\begin{align*}
	\E{B\,\mid\, A=\pm} {}={}& \frac{p(+, \pm) - p(-, \pm)}{p(+, \pm) + p(-, \pm)} \\
	{}={}& \frac{\E{B} \pm \E{AB}}{1 \pm \E{A}} \\
	{}={}& \frac{\hat{b}^T\vec\psi_1 \pm \hat{b}^T \Gamma \hat{a}}{1 \pm \vec\psi_2^T\hat{a}} \\
	{}={}& \hat{b}^T\frac{\vec\psi_1 \pm \Gamma\hat{a}}{1 \pm \vec\psi_2^T\hat{a}} \\
	{}\equiv{}& \hat{b}^T \left.\vec\psi_1\right|_{A=\pm}\,.
\end{align*}
We thus find that the conditional expectations for the measurement on qubit 1
are still linear in the axis $\hat{b}$, allowing us to identify the
\enquote{conditional state} $\left.\vec\psi_1\right|_{A=\pm}$ as in
\cref{eq:conditional}.

Using the parametrization from \cref{thm:bijective}, we check that
$\left.\vec\psi_1\right|_{A=\pm}$ is in fact a unit vector. To that end, using
the shorthand $\abs{\vec\psi_2} = \cos\rho$ as usual, note that
\begin{align*}
	\Gamma^T\Gamma {}={}& Q_{\psi_2}^T \bar{R}^T \bar{R} Q_{\psi_2} = Q_{\psi_2}^2 \\
	{}={}& \sin^2\rho + \vec\psi_2\otimes\vec\psi_2\left[ \frac{2\sin\rho}{1+\sin\rho} + \frac{\cos^2\rho}{\left(1+\sin\rho\right)^2} \right] \\
	{}={}& \sin^2\rho + \vec\psi_2\otimes\vec\psi_2\frac{2\sin\rho + 2\sin^2\rho + \cos^2\rho}{\left(1+\sin\rho\right)^2} \\
	{}={}& 1 - \abs{\vec\psi_2}^2 + \vec\psi_2\otimes\vec\psi_2\,.
\end{align*}
Then
\begin{align*}
	\abs{\vec\psi_1 \pm \Gamma\hat{a}}^2 {}={}& \abs{\vec\psi_1}^2 \pm 2\vec\psi_1^T\Gamma\hat{a} + \hat{a}^T\Gamma^T\Gamma\hat{a} \\
	{}={}& 1 \pm 2\vec\psi_2^T\hat{a} + \left(\vec\psi_2^T\hat{a}\right)^2 \\
	{}={}& \left( 1 \pm \vec\psi_2^T\hat{a} \right)^2\,,
\end{align*}
so indeed $\left.\vec\psi_1\right|_{A=\pm}$ as written in \cref{eq:conditional}
is normalized to unit length, as long as $1 + A\vec\psi_2^T\hat{a} \neq 0$.

The critical case $A\vec\psi_2^T\hat{a} = -1$ corresponds to a separable state
($\abs{\vec\psi_2} = 1$) in which we measure qubit 2 exactly along its Bloch
vector ($\hat{a}\parallel\vec\psi_2$) and obtain a \emph{negative} result $A =
-\E{A}$; which is impossible and thus irrelevant.

\bibliography{qubits}

\end{document}